%% file: main.tex
\makeatletter\def\input@path{{vendor-acmart/}}\makeatother
\documentclass[sigplan,10pt]{acmart}

\copyrightyear{2026}
\acmYear{2026}
\setcopyright{cc}
\setcctype{by}
\acmConference[EUROSYS '26]{21st European Conference on Computer Systems}{April 27--30, 2026}{Edinburgh, Scotland Uk}
\acmBooktitle{21st European Conference on Computer Systems (EUROSYS '26), April 27--30, 2026, Edinburgh, Scotland Uk}
\acmDOI{10.1145/3767295.3803603}
\acmISBN{979-8-4007-2212-7/2026/04}

\makeatletter
\newenvironment{tightcases}[1][1]{%
  \left\lbrace
  \def\arraystretch{#1}%
  \array{@{}l@{\quad}l@{}}%
}{\endarray\right.}
\makeatother

\title{HARP: Orchestrating Automated Parallel Training on Heterogeneous GPU Clusters} 

\input{sections/00-authors}

\input{sections/00-metadata}

\input{macros}  
\begin{document}

\renewcommand{\shortauthors}{Antian Liang et al.}

\begin{abstract}
With the rapid evolution of GPU architectures, the heterogeneity of model training infrastructures is steadily increasing. In such environments, effectively utilizing all available heterogeneous accelerators becomes critical for distributed model training. 
However, existing frameworks, which are primarily designed for homogeneous clusters, often exhibit significant resource underutilization when deployed on heterogeneous accelerators and networks.
In this paper, we present Harp, an automated parallel training framework designed specifically for heterogeneous clusters. Harp introduces a fine-grained planner that efficiently searches a wide space for the inter-operator parallel strategy, enabling Harp to alleviate communication overheads while maintaining balanced loads across heterogeneous accelerators.
In addition, Harp implements a heterogeneity-aware 1F1B scheduler that adaptively adjusts the execution timing and ordering of microbatches based on network characteristics, maximizing computation–communication overlap under cross-cluster interconnects while incurring only minimal memory overhead.
Our evaluation results show that Harp can deliver $1.3\times$–$1.6\times$ higher performance on heterogeneous clusters than state-of-the-art training frameworks. 
\end{abstract}

\maketitle

\begin{sloppypar} 

\input{sections/01-introduction}

\input{sections/02-background_and_motivations}
\input{sections/03-overview}

\input{sections/04-design1}

\input{sections/05-design2}
\input{sections/06-evaluation}

\input{sections/07-discussion}
\input{sections/08-related}

\input{sections/09-conclusion}



\balance

\bibliographystyle{vendor-acmart/ACM-Reference-Format}
\bibliography{refs}
\appendix
\input{EuroSys26_ArtifactAppendix_template}
\input{appendix/A-proof}
\input{appendix/B-MFU}

\input{appendix/C-HardwareSpec}
\input{appendix/D-evaluationMetis}

\end{sloppypar}
\end{document}

%% file: sections/00-authors.tex
\author{\texorpdfstring{Antian Liang$^{1,*}$, Zhigang Zhao$^{1,2,*}$, Kai Zhang$^{1,\dagger}$, Xuri Shi$^{1}$, Chuantao Li$^{2}$, Chunxiao Wang$^{2}$, Zhenying He$^{1}$, Yinan Jing$^{1}$, X. Sean Wang$^{1}$}{Antian Liang, Zhigang Zhao, Kai Zhang, Xuri Shi, Chuantao Li, Chunxiao Wang, Zhenying He, Yinan Jing, X. Sean Wang}}

\affiliation{%
  \institution{%
    \parbox{\linewidth}{\centering
      \vspace{2pt}
      $^1$Fudan University, Shanghai, China \\
      $^2$Shandong Computer Science Center (National Supercomputer Center in Jinan), Jinan, Shandong, China \\
      \vspace{2pt}
      \small $^*$Equal contribution, $^\dagger$Corresponding author.
    }
  }
  \country{} 
}

\email{{atliang23, zgzhao21}@m.fudan.edu.cn, zhangk@fudan.edu.cn, xrshi23@m.fudan.edu.cn}
\email{chuantaolisdsc@gmail.com, wangchx@sdas.org, {zhenying, jingyn, xywangCS}@fudan.edu.cn}

%% file: sections/00-metadata.tex
\ccsdesc[500]{Computing methodologies~Parallel computing methodologies}
\ccsdesc[500]{Computing methodologies~Machine learning}
\ccsdesc[500]{Computer systems organization~Cloud computing}

\keywords{Distributed Training, Heterogeneous GPU Clusters}

%% file: macros.tex
\usepackage{CJKutf8}
\usepackage{graphicx}  
\usepackage{float}  
\usepackage{subfigure}  
\usepackage{makecell}

\usepackage{booktabs}
\usepackage{multirow}

\usepackage{microtype}       

\usepackage{enumitem}

\usepackage[font=normalsize,skip=3pt]{caption}  
\usepackage{siunitx}
\usepackage[ruled,vlined,linesnumbered]{algorithm2e} 
\DontPrintSemicolon 


%% file: sections/01-introduction.tex
\section{Introduction}



In the pursuit of higher performance and broader applicability, the scale of deep learning models has rapidly grown to $O(10^{12}$–$10^{13})$ parameters in recent years (e.g., LLaMA-3 series, GPT models)\cite{gpt1, gpt2, gpt3, gpt4}. Meeting this demand requires massive device memory and computation resources, which model developers typically address by deploying large homogeneous clusters composed of identical accelerators.

Compared to large homogeneous clusters, heterogeneous clusters are often more accessible. 
First, due to transitional upgrades, organizations typically accumulate hardware over multiple years. With hardware vendors like NVIDIA releasing new architectures on an annual cycle, resource pools naturally become more heterogeneous over time. 
Second, high-end homogeneous resources are frequently scarce. In cloud environments, provisioning a large homogeneous cluster within a single availability zone often entails prohibitive wait times. For instance, it has been reported that acquiring 8 $\times$ A100 GPUs on-demand in a single zone on Google Cloud Platform (GCP) can require up to 7 hours of waiting. In contrast, provisioning 4 $\times$ A100 GPUs across two separate zones takes only approximately 2 hours on average~\cite{gcp, strati2025sailor}. This makes cross-zone, geo-distributed clusters significantly more accessible.


\begin{figure}[!t]
    \centering
    \includegraphics[width=1\linewidth]{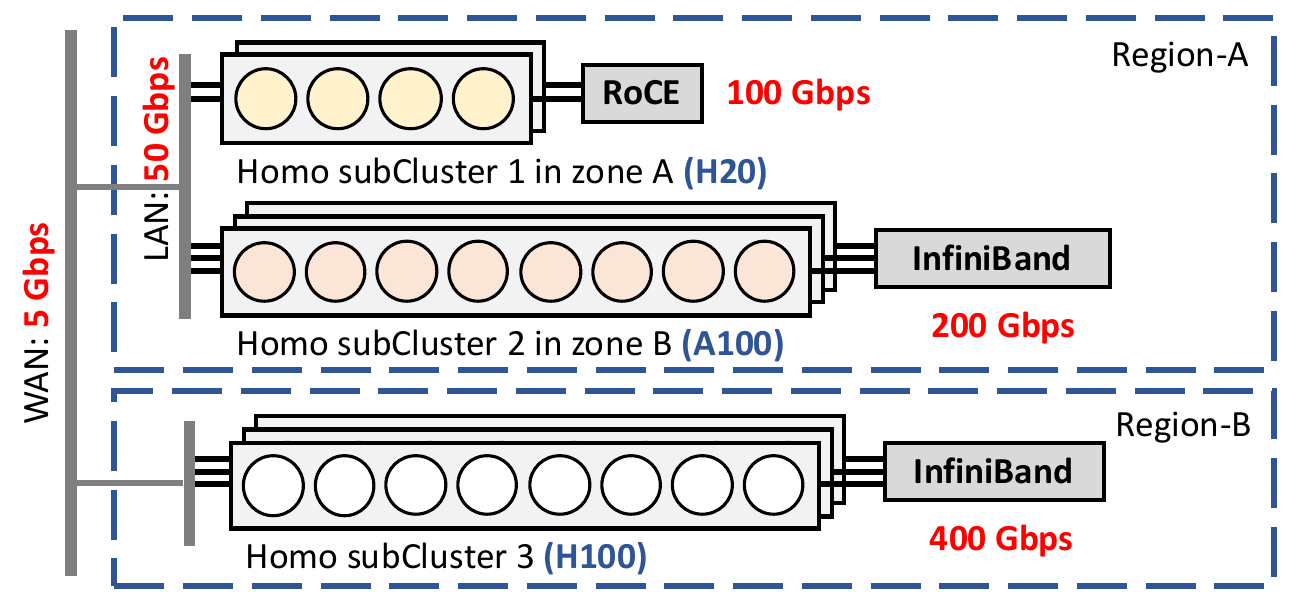}
    \caption{Example heterogeneous cluster composed of multiple homogeneous subclusters, with fast interconnects within subclusters but slower interconnects across them.}
    \label{fig:0}
\end{figure}

Model developers typically rely on distributed frameworks such as Megatron~\cite{Megatron-lm-v1, Megatron-lm-v2, Megatron-lm-v3}, DeepSpeed~\cite{DeepSpeed}, and Alpa~\cite{Alpa} for large-scale training. However, directly deploying these frameworks on dynamic, heterogeneous, and geo-distributed clusters leads to severe resource underutilization. Figure~\ref{fig:0} illustrates a representative heterogeneous cluster, where training tasks must contend with the challenges of computational heterogeneity and network heterogeneity. 
Regarding \textbf{computational heterogeneity}, frameworks for training on homogeneous clusters typically assume all accelerators have identical computation and memory capacities, thereby assigning equal workloads to them, causing significant load imbalancing. 
Regarding \textbf{network heterogeneity}, homogeneous subclusters distributed across different regions must often communicate over low-bandwidth interconnects (e.g., 1–10\,Gbps~\cite{um2024metis, DTFM, yan2024hexiscale}), whereas intra-subcluster communication typically enjoys high-bandwidth interconnects ranging from 10 to 800\,Gbps~\cite{SuperNIC}. This vast disparity in bandwidth causes inter-cluster communication to become a severe performance bottleneck, significantly slowing down the entire training process.
The combined impact of these computing and networking heterogeneity has been shown to drastically reduce training throughput in practice. For instance, Um et al.~\cite{um2024metis} reported that applying Alpa to train a Mixture-of-Experts (MoE) model on a mixed P100-V100 cluster resulted in up to a $7.4\times$ performance degradation.



To better exploit heterogeneous resources, 
recent studies~\cite{zhang2024poplar, um2024metis, yan2024hexiscale} have focused on the automated planning of heterogeneous parallel strategies by exploring the search space along intra-operator parallelism (e.g., data and tensor parallelism) and inter-operator parallelism (e.g., pipeline parallelism).
For example, Metis~\cite{um2024metis} and HexiScale~\cite{yan2024hexiscale} investigate:
(1) intra-operator (op) search space, where model or data tensors can be asymmetrically sliced so that heterogeneous accelerators executing the same operator in parallel are each assigned proportionally balanced workloads; 
and PipePar and Sailor~\cite{zhang2023pipepar} focus on
(2) inter-op search space, where the computation graph can be partitioned into pipeline stages with different workloads so that these stages are mapped onto accelerators with different capabilities, thereby achieving balanced execution across the pipeline.
While these systems achieve higher throughput than directly deploying frameworks originally designed for homogeneous clusters in heterogeneous environments, they still fail to reach the expected utilization (measured as Model FLOP Utilization, MFU) of heterogeneous clusters. For instance, as reported by Yan et al.~\cite{yan2024hexiscale}, training LLaMA-2 30B on a heterogeneous cluster yields only 17.1\% and 28.0\% MFU with Metis and HexiScale, respectively.

We analyze experimental results on heterogeneous clusters and observe that this under-utilization stems from three key factors.
\emph{\textbf{i)} Expensive cross-cluster intra-op communication.}
Exploring both intra- and inter-op search spaces can indeed improve load balance across heterogeneous accelerators. However, applying intra-op parallelism across subclusters connected by slow interconnects introduces substantial collective communication overhead~\cite{um2024metis, yan2024hexiscale, zhang2024hap}. In HexiScale’s experiments~\cite{yan2024hexiscale}, the overhead of communication is reported to be about 50\% of the total computation overhead.
\emph{\textbf{ii)} Limited inter-op search space.} 
The search space of inter-op parallel strategies scales with the number of operators in the computation graph, where a large-scale model typically consists of $10^4$–$10^5$ operators.
Exhaustively planning across this space is prohibitively expensive.
To reduce the cost, existing systems cluster operators into coarse-grained layers for planning~\cite{Alpa, yan2024hexiscale, strati2025sailor}. 
However, planning at a coarse layer granularity often fail to generate a balanced strategy on heterogeneous clusters, as it prevents precise alignment between pipeline stage workloads and the computational capacities of heterogeneous accelerators. For instance, experiments with PipePar~\cite{zhang2023pipepar} on training ResNet-50 using a heterogeneous GPU cluster showed that the best strategy still left the longest stage with 1.5$\times$ the compute cost of the shortest stage.
Planning at a finer granularity can narrow this gap, but the overhead grows rapidly as the granularity increases, quickly becoming prohibitive.
Our experiments show that training a GPT-39B model on 32V100–32A100 heterogeneous cluster can achieve balanced loads only at a granularity on the order of $10^2$ layers.
However, it takes more than five days to derive the parallel strategy with such a fine granularity, which is impractical.
\emph{\textbf{iii)} Cross-cluster inter-stage communication stalls computation.}
When inter-op parallelism spans multiple subclusters, limited cross-cluster bandwidth often causes inter-stage communication to block computation, leading to substantial pipeline bubbles. 
Although recent pipeline schedulers can partially hide this latency by launching additional forward microbatches during warmup, they are unaware of network heterogeneity and thus can hide only up to 50\% of the theoretical upper bound on inter-stage communication latency, while incurring substantial memory overhead from buffering nearly $2\times$ intermediate results~\cite{Alpa-opt}.

To bridge this gap, we introduce \textsc{Harp}, an automated framework for distributed parallel training deep learning models. Our key contributions are:

\noindent\textbf{(1) Fine-grained inter-op parallel strategy planner.}
To bypass prohibitive cross-cluster collective communication, \textsc{Harp} strategically confines intra-op parallelism within homogeneous subclusters and elevates inter-op planning to a significantly finer layer granularity than existing approaches to reclaim the lost search space. 
To overcome the planning overhead, we introduce a set of aggressive pruning and parallelization techniques. 
Specifically, We heuristically identify repeated modules to eliminate redundant profiling, preserving the computational graph’s structural hierarchy.
This design preserves the structural information in the original graph, which can then be exploited to extensively prune the profiling process. 
Building on this, we design a dynamic programming (DP) search algorithm to efficiently generate heterogeneous inter-op parallel strategies at fine layer granularity, enhanced with several system-level optimizations, including sparsity indexing, bidirectional pruning, and batched parallel search, which systematically accelerate the searching process.
These optimizations enable \textsc{Harp} to generate load-balanced strategies on fine layer granularity within minutes rather than days.

\noindent\textbf{(2) Heterogeneity-aware 1F1B scheduler.} 
\textsc{Harp} formulates communication–latency hiding as an optimization problem, optimizing the execution timing and ordering of microbatches at each stage. We model pipeline execution latency with a directed acyclic graph (DAG) abstraction and derive the H-1F1B scheduling strategy, which we formally prove to achieve the theoretical upper bound on the amount of communication latency that can be hidden, while incurring only minimal additional memory overhead.

\noindent\textbf{(3) Evaluations.} 
We implement \textsc{Harp}, evaluating it on clusters of up to 64 V100 and A100 GPUs. Across diverse heterogeneous settings, \textsc{Harp} achieves $1.3\times$–$1.6\times$ higher throughput than state-of-the-art baseline.
Moreover, on heterogeneous clusters where cross-cluster interconnects are limited to 5\,Gbps, \textsc{Harp} delivers throughput comparable to Megatron (latest version) running on homogeneous clusters fully connected via 200\,Gbps RDMA, under the same theoretical peak FLOP/s.
\footnote{We open-source \textsc{Harp} at \url{https://github.com/Lssyes/harp}.}

%% file: sections/02-background_and_motivations.tex
\section{Background and Motivations}
\label{sec:2}

\subsection{Background on Distributed Parallel Training}
Distributed training partitions the model and training data into smaller units for parallel execution across multiple accelerators. 
To achieve this, the system employs several parallel strategies, which typically includes intra-op parallelism (e.g., data~\cite{DeepSpeed} or tensor~\cite{Megatron-lm-v1} parallelism) and inter-op parallelism (e.g., pipeline parallelism~\cite{Gpipe}), and employs a scheduler that orchestrates how forward and backward computations are executed in parallel across the cluster~\cite{PipeDream, PipeDream-2BW, qi2023zero, li2021chimera}.

For parallel strategy planning, manual frameworks such as Megatron~\cite{Megatron-lm-v3} require developers to manually configure parallel strategies, which demands deep expertise in the underlying infrastructure and parallelization techniques. This burden has motivated the development of automated planners. 
For example, Alpa~\cite{Alpa} uses a hierarchical planner to automatically apply intra- and inter-op parallelism for homogeneous clusters. It linearizes the model’s computation graph into a topologically ordered operator sequence. The cluster is modeled as a \emph{DeviceMesh}$(N,M)$, with $N$ nodes and $M$ devices per node. The planner proceeds in two steps.
(i) Profiling: The operators are first clustered into $L$ layers to shrink the inter-op search space. The framework then enumerates all adjacent layer subsequence as candidate pipeline stages, and all possible submeshes $(n,m)$ sliced from the DeviceMesh $(N, M)$. For each candidate stage–mesh pair, the planner determines the optimal intra-op parallel strategy and profiles its execution information, including compute cost and memory footprint.
(ii) Searching: Based on the profiled results, the planner formulates the search for inter-op parallel strategies as an optimization problem that minimizes end-to-end latency while balancing compute cost across stages. The search algorithm compares the end-to-end latency for different combinations of stage–mesh pairs and selects the optimal one as the inter-op parallel strategy.

\begin{figure}
    \centering
    \includegraphics[width=1\linewidth]{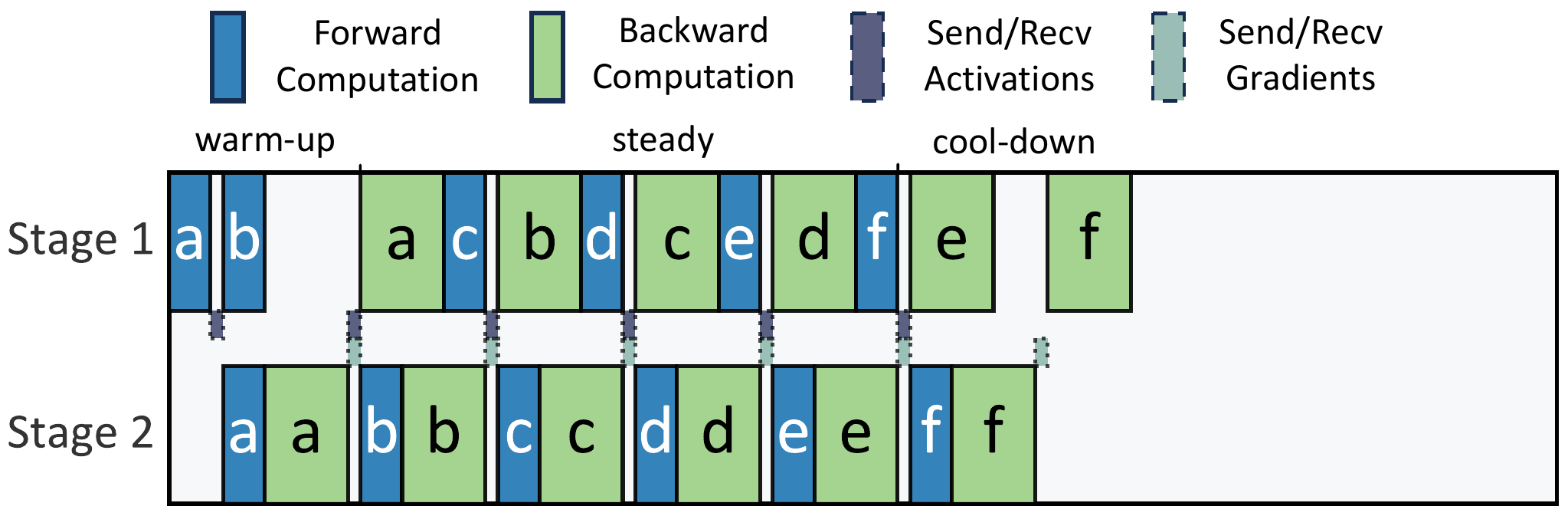}
    \caption{Classic 1F1B pipeline scheduler.}
    \label{fig:1f1b}
\end{figure}

For pipeline scheduling, it accelerates inter-op parallelism by splitting a batch into multiple microbatches that are executed in a pipelined manner~\cite{Gpipe, PipeDream, PipeDream-2BW}. 
Specifically, each microbatch completes its forward and backward passes by traversing all pipeline stages.
Once stage $i$ finishes the forward computation, its output activations are sent to stage $i{+}1$; after stage $i{+}1$ completes the backward computation, the resulting gradients are propagated back to stage $i$. 
The pipeline scheduler determines the timing and ordering of forward and backward computations across all stages.
As shown in Figure~\ref{fig:1f1b}, the classic 1F1B scheduler~\cite{PipeDream, PipeDream-2BW} proceeds in three phases:
(i) During the \emph{warm-up} phase, stage $i$ launches $(\#\text{stage}-i+1)$ forward microbatches (with $\#\text{stage}$ denoting the total number of stages);
(ii) in the \emph{steady} phase, each stage alternates between forward and backward passes in a one-forward–one-backward (1F1B) pattern;
(iii) in the \emph{cool-down} phase, the remaining backward passes are drained.

\subsection{Load Imbalance of Existing Parallel Strategy Planning on Heterogeneous Device}
\label{sec:2.2}
Planners designed for homogeneous clusters cannot adapt in heterogeneous settings, as they assume identical devices and assign equal workloads, leading to severe load imbalance and low utilization~\cite{um2024metis, yan2024hexiscale}.

\subsubsection{Limited Search Space Incurs Load Imbalance}
Recent work~\cite{zhang2024hap, um2024metis, yan2024hexiscale, li2025hetu, xu2024hethub}
has begun to address load imbalance in heterogeneous clusters by introducing heterogeneity-aware parallelization strategies across both inter-op and intra-op parallelism.
For instance, Poplar allocates different minibatch sizes to GPUs according to their compute capabilities and memory capacities to balance workloads in data parallelism~\cite{poplar}. 
Metis further extends heterogeneous-aware planning by exploring parallel strategy search across both inter-op and intra-op parallelism~\cite{um2024metis}. 
HexiScale advances this direction by enabling asymmetric tensor sharding across mixed GPU types, allowing different devices to process uneven tensor partitions~\cite{yan2024hexiscale}.
However, intra-op parallelism requires collective communication across all participating devices~\cite{Megatron-lm-v1, GSPMD, li2025hetu}. In many real-world heterogeneous clusters, cross-cluster communication is limited to slow Ethernet or even WAN links. Such slow networks significantly slow down collective communication.
As reported in the experiments of Yan et al.\cite{yan2024hexiscale}, when the inter-cluster bandwidth is limited to 5 Gbps, training LLaMA-2 70B with Metis and HexiScale achieves only 17.1\% and 27.2\% MFU, respectively.

Another approach confines intra-op parallelism to homogeneous meshes and introduces heterogeneity only at the inter-op level~\cite{Swarm-pp, zhang2023pipepar}. This avoids cross-cluster collective communication over slow links, but restricts the search space of candidate meshes. At the same time, inter-op strategy planning is usually performed at a coarse layer granularity to avoid high profiling and searching overhead, which limits the search space of candidate stages~\cite{Alpa, um2024metis,yan2024hexiscale}. Together, these two restrictions yield a much smaller search space of candidate stage–mesh pairs, making it difficult to balance compute costs across stages.
For example, under this restricted search space, experiments with \emph{PipePar} on ResNet-50 using a $(4\times2)$ heterogeneous GPU cluster showed that the best strategy left the longest stage with 1.5$\times$ the compute cost of the shortest~\cite{zhang2023pipepar}. While planning at finer layer granularity can make a more balanced parallel strategy, however, profiling scales as $\sim O(L^2)$ and search as $\sim O(L^5)$. According to Alpa’s experiments~\cite{Alpa}, even on homogeneous clusters, generating a parallel strategy with \#$L=8$ layers for GPT-39B already takes about an hour~\cite{Alpa}. We evaluate planning overhead at finer granularity(e.g., \#$L=146$), and it increases dramatically, which can take more than 5 days, even erasing the training-time saved by a more balanced parallel strategy.

\begin{figure}
    \centering
    \includegraphics[width=1\linewidth]{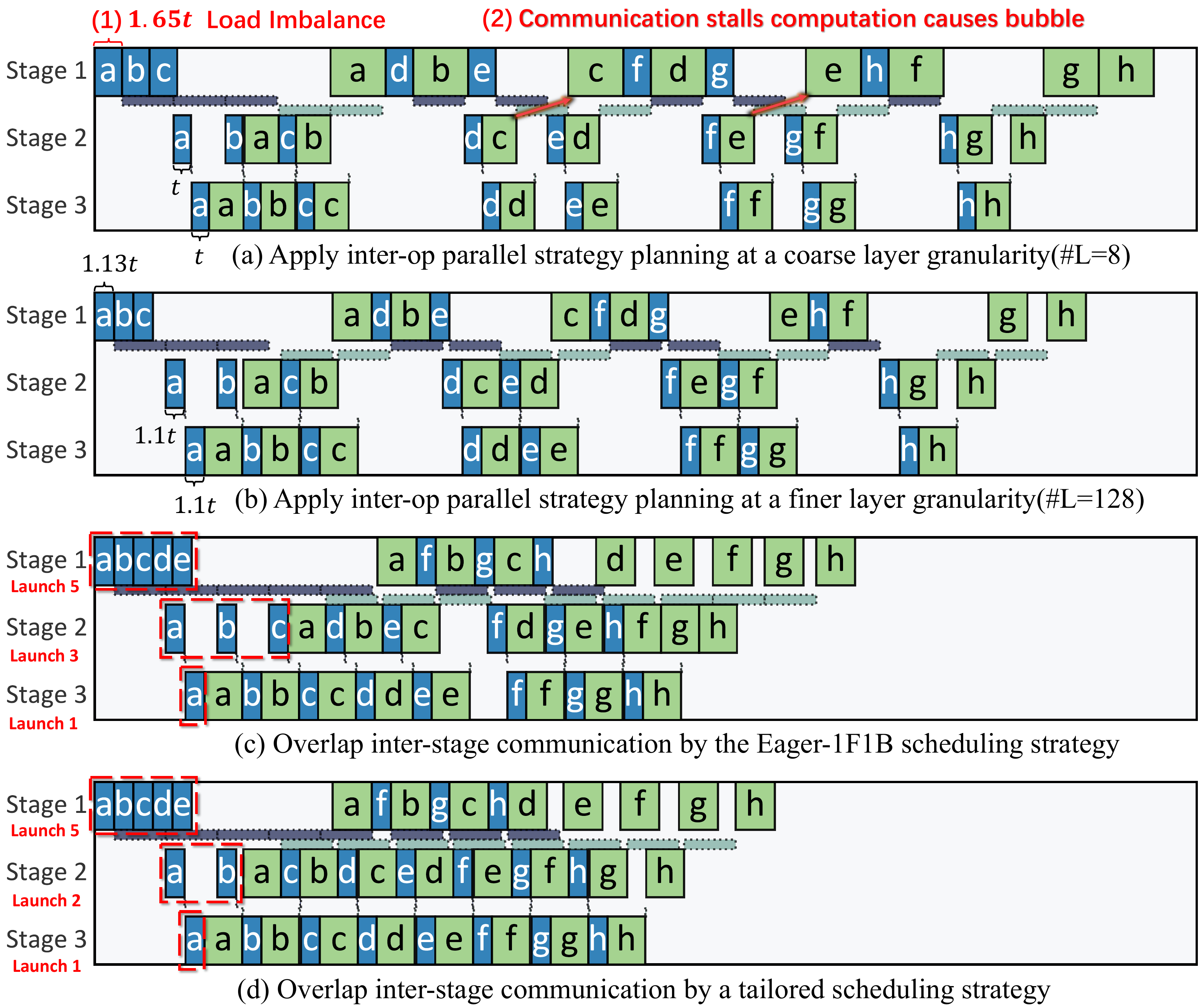}
    \caption{Pipeline timeline of case studies.}
    \label{fig:2:casestudy}
\end{figure}

\subsubsection{Case Study}
We conduct a case study to examine how different layer granularities in inter-op parallelism affect load balance, under the constraint that intra-op parallelism is restricted to homogeneous meshes.

We train a GPT model on a heterogeneous cluster consisting of two meshes: $DeviceMesh_{A100}(2,2)$ with 2$\times$2 A100 GPUs, and $DeviceMesh_{V100}(1,2)$ with 1$\times$2 V100 GPUs. The A100 nodes are connected by a 200\,Gbps InfiniBand network, while cross-cluster communication is limited to a 5 Gbps Ethernet link. We use an off-the-shelf operator clustering tool\cite{alpa-github} to group the layer sequence into 8 (same with Alpa) and 128 layers, respectively, with each layer having approximately equal workload. We explore the inter-op parallel strategies in this constrained search space and summarize the optimal result in Table \ref{tab:1}.

As shown in Figure~\ref{fig:2:casestudy}(a), inter-op parallel strategy planning at a coarse layer granularity (\#$L=8$) leads to severe load imbalance. Layer 1–2 are mapped to stage 1 running on 2× V100 GPUs, where only 16.8\% of the total compute capacity is responsible for 25\% of the workload. Using the compute cost of a forward microbatch in stages 2 and 3 as a reference $t$, stage 1 requires $1.65t$ to complete its computation. This imbalance arises because the restricted search space cannot adequately match workloads to the heterogeneous compute capabilities. In contrast, Figure~\ref{fig:2:casestudy}(b) shows planning at a finer granularity (\#$L=128$). Here, stage 1 is assigned less workload, while stages 2 and 3 are assigned more, leading to a computation graph partitioning that more closely matches each mesh’s compute capacity. This fine-grained plan achieves improved load balance and higher throughput. When $B=128$ microbatches, assuming full inter-stage communication overlap, the fine-grained plan achieves a 40.1\% throughput improvement over the coarse-grained plan.

\begin{figure*}[!t]
    \centering
    \includegraphics[width=1\linewidth]{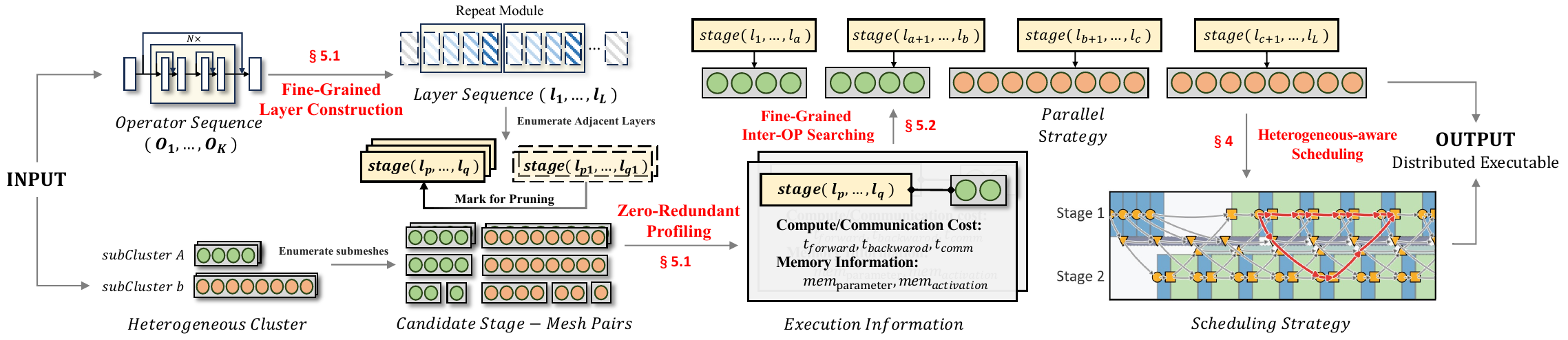}
    \caption{Overview of \textsc{Harp} workflow.}
    \label{fig:3:overview}
\end{figure*}

\begin{table}[t]
\centering
\caption{Inter-op parallel strategies for the case study. Notation: $mesh_{V100}(1,2)$ denotes a homogeneous submesh with ($1\times2$)V100 GPUs; the [16.8\%] indicates this submesh’s share of the cluster’s peak FLOP/s.}\setlength{\tabcolsep}{4pt}     
\renewcommand{\arraystretch}{1.05}
\small
\begin{tabular}{c c c c c c}
\toprule
\textbf{\#L} & \textbf{Stage} & \textbf{Submesh} & \textbf{Layers} & \textbf{Layer\%} & \textbf{Cost} \\
\midrule
\multirow{3}{*}{}
 & 1 & \makecell{$mesh_{V100}(1,2)${\scriptsize[16.8\%]}} & 1,2     & 25.0\% & $1.65t$ \\
 8& 2 & \makecell{$mesh_{A100}(1,2)${\scriptsize[41.6\%]}} & 3,4,5   & 37.5\% & $t$     \\
 & 3 & \makecell{$mesh_{A100}(1,2)${\scriptsize[41.6\%]}} & 6,7,8   & 37.5\% & $t$     \\
\midrule
\multirow{3}{*}{}
 & 1 & \makecell{$mesh_{V100}(1,2)${\scriptsize[16.8\%]}} & 1--22   & 17.2\% & $1.13t$ \\
 128& 2 & \makecell{$mesh_{A100}(1,2)${\scriptsize[41.6\%]}} & 23--76  & 41.4\% & $1.10t$ \\
 & 3 & \makecell{$mesh_{A100}(1,2)${\scriptsize[41.6\%]}} & 76--128 & 41.4\% & $1.10t$ \\
\bottomrule
\label{tab:1}
\end{tabular}
\end{table}

\subsection{Limited Latency Hiding of Pipeline Schedulers on Heterogeneous Networks}
In the previous case study, we observed that communication stalls are another significant source of performance degradation, as the classic 1F1B scheduler enforces strict data dependencies, often forcing computation to wait for slow inter-stage communication, which creates pipeline bubbles during the steady phase.


\subsubsection{Overlap-friendly Scheduling Strategy}
Communication stalls in the steady phase can be mitigated by overlapping communication with computation. For example, \emph{Eager-1F1B}~\cite{Alpa-opt}, a variant of the classic 1F1B scheduler, launches $2\times(\#\text{stage}-i)+1$ forward microbatches during warm-up at stage~$i$. Concretely, this means that stage~$i$ launches two more forward microbatches than stage~$i{+}1$, thereby creating more opportunities to hide communication latency.
However, it applies a fixed scheduling strategy, where each stage launches a static number of extra forward microbatches, which is less effective under heterogeneous networks. Specifically:
(i) even when inter-stage communication latency is small and requires no hiding, the scheduler still launches two additional forward microbatches. For large-scale models with 16 or more stages, this behavior nearly doubles the activation memory compared to the classic 1F1B; and
(ii) its ability to hide communication is fundamentally capped at 50\% of the theoretical upper bound. Let $t_f$ and $t_b$ denote the forward and backward compute cost of a stage. With Eager-1F1B, only inter-stage communication whose cost does not exceed $(t_f{+}t_b)/2$ can be fully overlapped, which represents 50\% of the theoretical upper bound.

\subsubsection{Case Study}
We conduct a case study to illustrate why existing overlap-friendly schedulers perform suboptimally on heterogeneous slow networks and to motivate our tailored scheduling strategy. The setup uses the same model, cluster, and parallel strategy as in Figure~\ref{fig:2:casestudy}(b), but applies different pipeline scheduling strategies.

As shown in Figure~\ref{fig:2:casestudy}(c), Eager-1F1B launches $\{5,3,1\}$ forward microbatches at stages 1–3. In this strategy, stage~2 launches two more microbatches than stage~3, even though the two are connected via InfiniBand and incur negligible communication latency. Overlapping such fast communication is unnecessary and only wastes memory.
Similarly, stage~1 launches two more microbatches than stage~2, which is still insufficient to fully hide the communication latency between them. Such delays over slow interconnects are frequently encountered in heterogeneous environments\cite{DTFM, wu2024atom}.

To further eliminate the steady-phase bubbles observed in Fig.~\ref{fig:2:casestudy}(c), we manually configure a tailored scheduling strategy, illustrated in Fig.~\ref{fig:2:casestudy}(d). Compared to Eager-1F1B, this strategy launches $\{5,2,1\}$ forward microbatches at stages 1–3, respectively. It avoids redundant launches between stages~2 and~3 which are connected by fast link, while increasing the communication-hiding capacity between stages~1 and~2 where communication cost range from $(t_f{+}t_b)/2 $ to $(t_f{+}t_B)$. In this way, additional launches occur only when necessary. For large-scale models with 16 or more stages on heterogeneous networks, this strategy significantly reduces activation memory consumption compared to Eager-1F1B, while achieving greater communication overlap.

%% file: sections/03-overview.tex
\section{Overview}

\label{section:3overview}
To address these challenges, we propose \textsc{Harp}, a system that automates parallel training of deep learning models on heterogeneous GPU clusters. Specifically, to address load imbalance, \textsc{Harp} apply inter-op parallelism by partitioning the computation graph on fine-grained layers and map them onto a heterogeneous cluster, while retaining standard intra-op parallelism within homogeneous clusters. 
This design avoids heavy collective communication overheads that would otherwise occur on cross-cluster links. 
To resolve the trade-off between load imbalance (caused by coarse layer granularity) and inter-op parallel strategy planning overhead (caused by fine layer granularity),
\textsc{Harp} iteratively scans the computation graph, heuristically identifies repeated and non-repeated modules, and clusters operators within each module into layers, concatenating to a layer sequence that preserves the repeated structural patterns of the original graph. These structural layers generate multiple identical candidate stage–mesh pairs, which is exploited by our profiler to perform efficient pruning, achieving both fine layer granularity and low profiling overhead.
Based on the profiling results, we formulate inter-op parallel strategy searching as an optimization problem and design a DP algorithm to derive balanced parallel strategies across stages for the heterogeneous cluster. Furthermore, we incorporate pruning and system-level optimizations to alleviate the search overhead on such fine-grained layers.
In addition, to mitigate communication stalls in heterogeneous networks, we formulate the communication–latency hiding problem in pipeline scheduling as an optimization problem which resolves the number of forward microbatches to launch during warm-up for each stage. Building on this formulation, we model the end-to-end latency using a DAG abstraction and adaptively derive an optimal scheduling strategy that maximizes latency hiding while incurring only minimal additional memory overhead.


To achieve these, as shown in Figure~\ref{fig:3:overview},
\textsc{Harp} takes a topologically ordered operator sequence and a heterogeneous cluster as input. 
It first performs \emph{fine-grained layer construction} (§\ref{sec:5.1}) on the operator sequence and output structural layers.
Next, \textsc{Harp} slices the heterogeneous cluster into submeshes and enumerates all adjacent layer subsequences as candidate stages. 
Since many of these stage candidates are structurally identical, \textsc{Harp} prunes the redundant stages and performs \emph{zero-redundant profiling} (§\ref{sec:5.1}) to collect execution information for each unique stage–mesh pair, including compute cost, communication cost, and memory footprint.
Based on these, we perform \emph{fine-grained inter-op searching} (§\ref{sec:5.2}) to generate parallel strategies efficiently, applying several system-level optimizations including sparsity indexing, bidirectional pruning, and batched parallel search to systematically accelerate strategy generation.
The H-1F1B scheduler then generates a \emph{heterogeneity-aware scheduling} (§\ref{section:4}) strategy that based on network characteristics to maximize computation–communication overlap on slow links.
Finally, the combined scheduling and parallel strategies are compiled into a distributed executable for model training.

%% file: sections/04-design1.tex
\section{Heterogeneity-aware 1F1B Scheduler}
\label{section:4}
In this section, we first introduce the design of the H-1F1B scheduler, as inter-op parallel strategy planning depends on the memory constraints and end-to-end cost model imposed by the pipeline scheduler.

\subsection{Design of H-1F1B}
\begin{figure}
    \centering
    \includegraphics[width=1\linewidth]{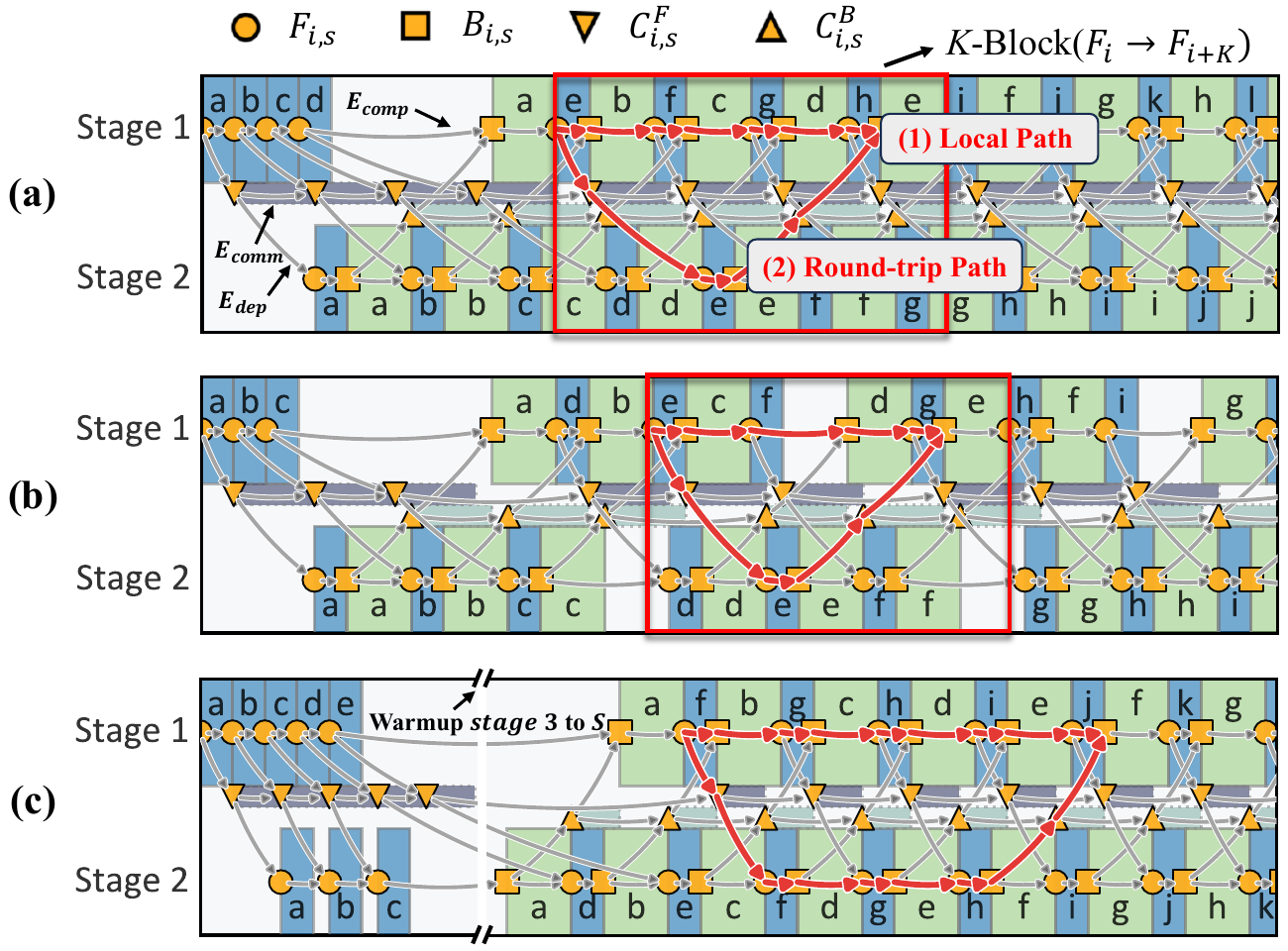}
    \caption{DAG representation of the pipeline execution. 
    (a) Two-stage homogeneous case where the local path dominates $K$-block latency. 
    (b) Two-stage homogeneous case where the round-trip path dominates $K$-block latency. 
    (c) $S$-stage homogeneous case, reduced to an equivalent two-stage subgraph by pruning nodes \& edges unrelated to stages~1–2.}

    \label{fig:pipeline_dag}
\end{figure}

Unlike 1F1B~\cite{PipeDream} and Eager-1F1B~\cite{Alpa-opt}, which launch a fixed number of forward microbatches at each stage during warm-up, H-1F1B adopts a tailored strategy that adjusts the launch count based on inter-stage communication costs.
Specifically, proceeding in descending stage order from $S$ to~1, each stage $i$ launches one more forward microbatch than stage $i{+}1$.
When the communication cost between stage~$i$ and stage~$i{+}1$ is higher, stage~$i$ can launch additional forward microbatches to better overlap communication.
The intuition is that launching more forward microbatches enlarges the temporal gap between the forward and backward passes of the same microbatch in the steady phase, creating more slack to hide communication latency. 
For example, as shown in Figure~\ref{fig:pipeline_dag}(a,b), although both cases incur the same compute and communication cost per stage, launching four forward microbatches creates a larger gap than three, thereby enabling more inter-stage communication to be overlapped.
Thus, larger communication costs are compensated by proportionally larger launch counts.
Let $N_i$ denote the number of forward microbatches launched by stage~$i$ during warm-up. When deployed in a heterogeneous cluster, H-1F1B adaptively derives $N_i$ as follows and the derived $N_i$ can achieve the theoretical upper bound in hiding the network latency, which will be formally proved in § 4.2.

Considering a pipeline with $S$ stages under a given parallel strategy. Let $t_i$ denote the per-microbatch computation cost (forward plus backward) of stage~$i$, and let $c_i$ denote the per-microbatch communication cost of transferring activations or gradients between stage~$i$ and stage~$i{+}1$ (with $c_S=0$). Define $t_{\max} = \max_{i \in \{1,\dots,S\}} t_i$ as the maximum per-stage computation cost. We assume $c_i \leq t_{\max}$ for all $i$; otherwise, full communication–computation overlap would be impossible. The last stage always launches a single microbatch ($N_S=1$). For all $i \in \{1,\dots,S{-}1\}$, we define
\begin{equation}
N_i = 1 + \sum_{k=i}^{S-1} \delta_k,
\label{eq:Ni}
\end{equation}
where $\delta_i$ denotes the number of additional forward microbatches launched by stage~$i$ relative to stage~$i{+}1$. The value of $\delta_i$ is determined by the ratio of $c_i$ to $t_{\max}$:
\begin{equation}
\delta_i =
\begin{tightcases}
1, & 0 < c_i \le \varepsilon \cdot t_{\max},\\
2, & \varepsilon \cdot t_{\max} < c_i \le t_{\max}/2,\\
3, & t_{\max}/2 < c_i \le t_{\max}
\end{tightcases}
\label{eq:delta_i}
\end{equation}
where $\varepsilon$ is a small positive constant.

\subsection{Proof of Effectiveness and Generality}
We formally prove that H-1F1B achieves the optimal communication–computation overlap, starting with a formal representation of pipeline execution.

\subsubsection{DAG Formulation}
We abstract the execution of 1F1B-style pipelines as a DAG, defined as follows.

\noindent\textbf{Nodes.} As shown in Figure~\ref{fig:pipeline_dag}, for each microbatch $i$ at stage $s$, the DAG includes four types of nodes:
\boldmath$F_{i,s}$\unboldmath\ (forward compute),
\boldmath$B_{i,s}$\unboldmath\ (backward compute),
\boldmath$C^F_{i,s}$\unboldmath\ (forward communication from stage $s$ to $s{+}1$), and
\boldmath$C^B_{i,s}$\unboldmath\ (backward communication from stage $s{+}1$ to $s$).
We also define a \emph{sink node} that connects to all terminal nodes via zero-weight edges, ensuring the DAG is fully connected.

\noindent\textbf{Edges.} The DAG includes three categories of edges. 
i) \boldmath$E_{\mathrm{comp}}$\unboldmath: enforce sequential execution of computations within the same stage;
ii) \boldmath$E_{\mathrm{comm}}$\unboldmath: enforce sequential transmissions in the same direction over a link (assuming full-duplex interconnects);  
iii) \boldmath$E_{\mathrm{dep}}$\unboldmath: capture the ordering between computation and communication within each microbatch.

\noindent\textbf{Durations.} Each edge $(u \to v)$ is weighted by the execution cost of node $u$, enforcing the standard finish-to-start precedence constraint:
\begin{equation}
s(v) - s(u) \ge d(u), \label{eq:precedence}
\end{equation}
where $s(u)$ is the start time of node $u$ and $d(u)$ its execution cost\footnote{If activation recomputation is used, the overhead is included in the corresponding backward compute node.}.
We initialize the schedule with $s(F_{1,1}) = 0$.

\subsubsection{Two-stage Balanced Pipeline Case ($S{=}2$)}
\label{sec:4.2.2}
We begin the proof with the base case of a two-stage pipeline where both stages have identical per-microbatch compute costs. The argument will then be extended to pipelines with $S$ stages via mathematical induction, and finally generalized to uneven per-stage compute costs.

Let the compute costs of stage $s \in \{1,2\}$ be $d(F_{i,s}) = f$ and $d(B_{i,s}) = b$, with symmetric communication such that $d(C^F_{i,1}) = d(C^B_{i,1}) = c$. During warm-up, stage~1 launches $K = 1 + \delta_1$ forward microbatches, while stage~2 launches only one. The steady phase dominates the overall latency of a 1F1B-style pipeline, with its duration being at least $(B-K)(f+b)$. The minimum value is attained when all inter-stage communications are fully overlapped with computation, i.e., when no pipeline bubbles remain in the steady phase. This lower bound also determines the theoretical upper bound of communication cost that can be hidden by overlap. We therefore restrict our analysis to
\begin{equation}
0 \le c \le f{+}b, \label{eq:overlap-regime}
\end{equation}
since full communication–computation overlap is impossible when outside this range.  
Given this condition, the key question becomes: \emph{what is the minimum $K$ that minimizes the steady-phase latency?} 
To answer this, we first derive an expression for the steady-phase latency as a function of $K$. 
Specifically, we begin by analyzing the temporal distance between $F_{i,1}$ and $B_{i,1}$ of the same microbatch.

\noindent\textbf{Lemma 1: Precedence Bound within a Microbatch.} 
For any microbatch $i$ at stage~1 in the steady phase, 
\begin{equation}
s(B_{i,1}) = s(F_{i,1}) + \max \big\{K\cdot f+(K{-}1)\, b,\, 2f+b+2c\big\}. \label{eq:lemma1}
\end{equation}

\emph{Proof.}
Between $F_{i,1}$ and $B_{i,1}$, there are two possible paths, shown in Fig.~\ref{fig:pipeline_dag}(a).  
\textbf{(1) Local path.} With $K$ forward microbatches launched in warm-up, stage~1 accumulates a backlog of $K$ forward and $(K{-}1)$ backward computations ahead of $B_{i,1}$ in the steady phase. Serializing these nodes gives a path length of $Kf+(K{-}1)b$.  
\textbf{(2) Round-trip path.} To obtain the gradient required by $B_{i,1}$, microbatch $i$ must traverse the path $F_{i,1}\!\rightarrow\!C^F_{i,1}\!\rightarrow\!F_{i,2}\!\rightarrow\!B_{i,2}\!\rightarrow\!C^B_{i,1}$, whose total length is $2f+b+2c$.  
Since every path from $F_{i,1}$ to $B_{i,1}$ must satisfy the precedence constraint in Eq.~\eqref{eq:precedence}, we obtain
\begin{equation}
s(B_{i,1}) \;\ge\; s(F_{i,1}) \;+\; \max\!\big\{K \cdot f+(K{-}1)b,\; 2f+b+2c\big\}.
\label{eq:great}
\end{equation}

Under an efficient pipeline execution, any computation or communication node is scheduled to start immediately upon the completion of all its predecessor nodes to avoid unnecessary idle time. Consequently, the start time $s(B_{i,1})$ must attain the lower bound specified in Eq.~\eqref{eq:great}. This completes the proof of Lemma 1.

\noindent\textbf{Lemma 2: $K$-Block.} 
In the steady phase, we define a \emph{$K$-block} on stage~1 as the execution segment that begins at $F_{i,1}$ and ends at $F_{i+K,1}$. The start times of these nodes satisfy
\begin{equation}
s(F_{i+K,1}) \;=\; s(F_{i,1}) \;+\; \max\{\,K(f+b),\;\;2(f+b+c)\,\}\,.
\label{eq:kblock}
\end{equation}

\emph{Proof.}
On stage~1, $F_{i+K,1}$ can only begin after $B_{i,1}$ finishes, so $s(F_{i+K,1}) - s(B_{i,1}) = b$.  
Substituting the bound on $s(B_{i,1})$ from Eq.~\eqref{eq:lemma1} gives Eq.~\eqref{eq:kblock}.  
Thus, Lemma~2 holds.

\noindent\textbf{End-to-End Latency Analysis.}
For $B$ microbatches, the steady phase can be partitioned into $\lfloor (B{-K})/K \rfloor$ disjoint $K$-blocks, each of duration
\begin{equation}
\Lambda_K = s(F_{i+K,1}) - s(F_{i,1})= \max\{K(f+b),\,2(f+b+c)\}.
\end{equation}
Hence, when $B$ is sufficiently large,  the end-to-end latency is approximated by
\begin{equation}
T(K) \;\approx\; \tfrac{B}{K}\cdot \Lambda_K \;+\; O(1) \;\approx\; B \cdot \max\!\Big\{f+b,\; \tfrac{2(f+b+c)}{K}\Big\}\,,
\end{equation}
where $O(1)$ denotes lower-order overhead.

To minimize $T(K)$, $K$ must be large enough so that $2(f+b+c)/K \le f+b$.  
Equivalently, the required number of extra forward launches can be expressed as 
\begin{equation}
\delta_1 \;=\; K - 1 \;=\; \Big\lceil 1 + \tfrac{2c}{f+b} \Big\rceil.
\label{eq:result2h}
\end{equation}
Theoretically, achieving a strictly bubble-free steady phase requires $\delta_i = \lceil 1 + 2c_i/t_{\max} \rceil \ge 2$ for any $c_i > 0$, as formally proved in Eq.\eqref{eq:result2h}. However, to prevent excessive memory inflation when communication overhead is marginal, H-1F1B introduces a tolerance threshold $\varepsilon$ and adopts a practical scheduling strategy as defined in Eq. \eqref{eq:delta_i}. This strategy balances the communication-computation overlap with activation memory constraints in heterogeneous environments.

\subsubsection{S-stage Balanced Pipeline Case.}
We now extend the result to a pipeline with $S$ stages via mathematical induction.  
Assume that for a pipeline with $(S{-}1)$ stages, the extra forward launches $\delta_i$ at stage~$i$ follow the H-1F1B rule:
\begin{equation}
\delta_i \;=\; N_i-N_{i+1}\;=\;\Big\lceil 1 + \tfrac{2c_i}{f+b} \Big\rceil, 
\quad \forall i \in \{1,\dots,S{-}1\},
\label{eq:resultsh}
\end{equation}
which guarantees minimum steady-phase latency and bubble-free execution.
Now consider a pipeline with $S$ stages, where stages~$2$ through $S$ are scheduled according to the $(S{-}1)$-stage case, leaving only stage~1. The remaining question is: \emph{what is the minimum $\delta_1$ achieves the same guarantee?}

The key observation is that  the start times of stage~3 through stage~$S$ are determined recursively by the start time of stage~2. Thus, as long as the relative start times of stage~2 nodes (with respect to $F_{1,2}$) remain unchanged, the schedules of all later stages are also unaffected.
As illustrated in Fig.\ref{fig:pipeline_dag}(c), this allows us to prune all nodes and edges unrelated to stages~1 and 2, reducing the analysis to an equivalent two-stage subgraph.

In this reduced setting, the steady-phase latency can be analytically formulated as a function of the launch count $N_1$. While the underlying logic parallels the two-stage case discussed in §\ref{sec:4.2.2}, the formulation involves a more complex generalization of dependency paths across all stages. The detailed derivation is provided in Appendix \ref{Appendix:A}. 
By optimizing this formulation, we find that minimizing the end-to-end latency requires $N_1$ to satisfy the condition in Eq.~\eqref{eq:resultsh}. Thus, by induction, the H-1F1B scheduler achieves the theoretical minimum steady-phase latency and guarantees bubble-free execution for pipelines with an arbitrary number of stages.

\subsubsection{S-stage Imbalanced Pipeline Case.}
To extend our analysis to imbalanced pipelines with varying per-stage computation costs, we conceptually "stretch" the computation time of all faster stages to match that of the bottleneck stage ($t_{\max}$). Through this reduction, an imbalanced pipeline is transformed into an balanced one, allowing the theoretical conclusions derived in the previous sections to be directly applied. Furthermore, since our fine-grained inter-op planner is designed to generate partitions with near-uniform computational costs, the deviation between individual stages and $t_{\max}$ is inherently minimal. Consequently, the analytical overhead introduced by this conservative approximation remains negligible. This ensures that the H-1F1B scheduler remains both effective and robust in practical heterogeneous environments without introducing significant performance discrepancies.

%% file: sections/05-design2.tex
\section{Fine-grained Heterogeneous Inter-op Parallel Strategy Planner}
\label{section:5}

\subsection{Zero-Redundant Profiler}
\label{sec:5.1}

\begin{figure}[t]
    \centering
    \includegraphics[width=\linewidth]{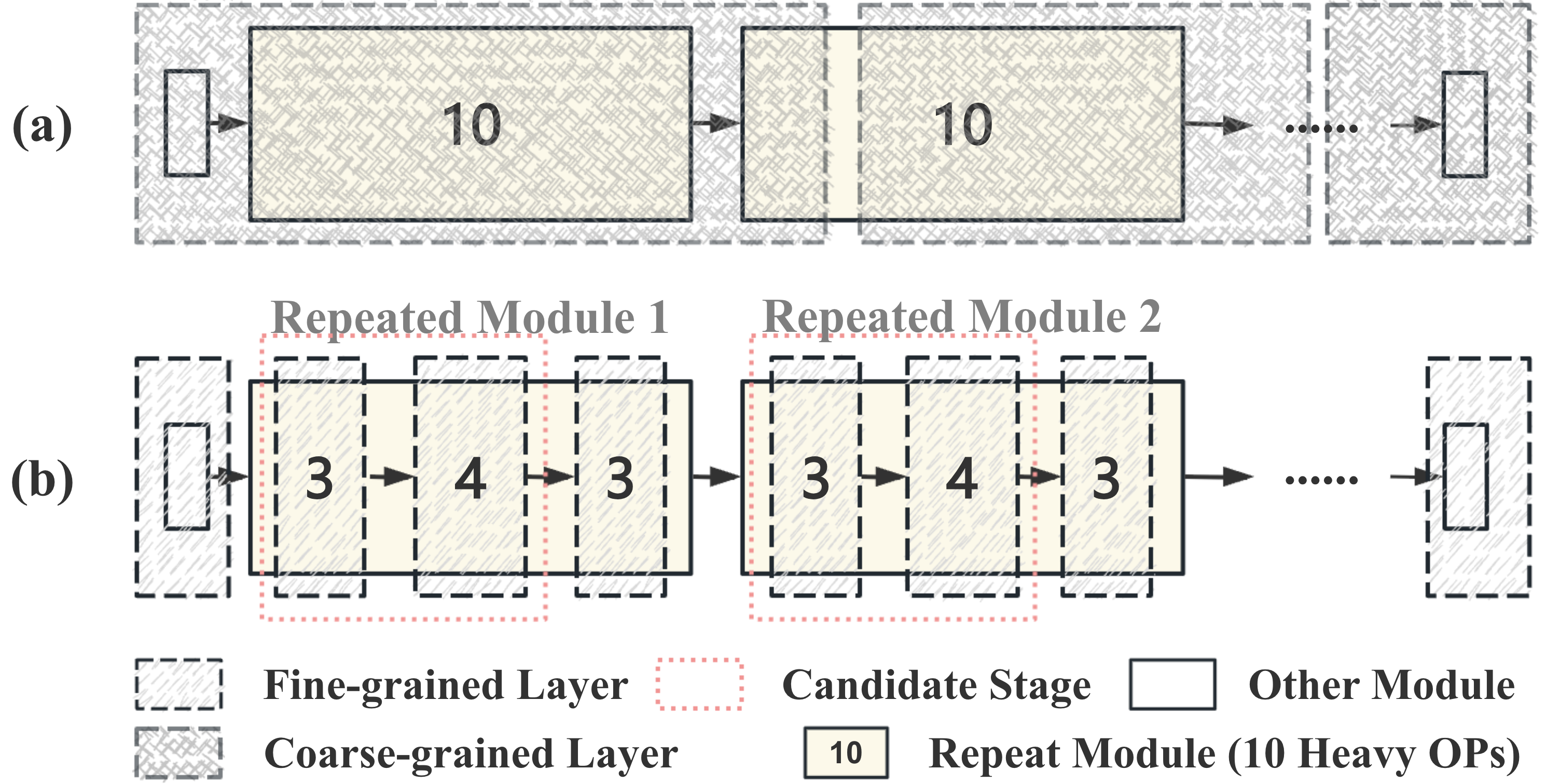}
    \caption{(a) Behaviour of layer construction in Alpa.  
             (b) Behaviour of layer construction in Zero-Redundant Profiler.}
    \label{fig:profiler}
\end{figure}

Before planning inter-op parallel strategies, the topologically ordered operator sequence is typically clustered into coarse-grained layers to reduce the profiling and search overhead.
As illustrated in Figure~\ref{fig:profiler}(a), existing methods simply cluster operators by compute cost, aiming to balance workload across layers~\cite{Alpa, um2024metis}.
However, this discards the model’s structural information, which limits opportunities for pruning during profiling. Consequently, planners are restricted to coarse-layer granularity to keep profiling time practical.

To address this limitation, we propose the Zero-Redundant Profiler. 
First, it constructs a fine-grained layer sequence while preserving structural information from the model.
As illustrated in Figure~\ref{fig:profiler}(b), this is done by partitioning the operator sequence into \emph{repeated modules} and \emph{non-repeated modules} through an iterative procedure: we initially treat the entire operator sequence as a single non-repeated module, and then repeatedly perform the following steps:
\textbf{(i)} among all current non-repeated modules, identify the most frequent contiguous sub-sequence that contains at least $z$ compute-intensive operators (e.g., MatMul or Convolution), where $z$ is a hyperparameter that controls the granularity of layer construction.
\textbf{(ii)} if such a sub-sequence exists, designate it as a repeated module and continue with step (i); otherwise, terminate.
For each module, we independently cluster its operators using existing algorithms~\cite{Alpa}, and then concatenate the results across modules to form a fine-grained layer sequence.

Next, once the layers are constructed, the profiler enumerates all candidate stage–mesh pairs and applies two forms of pruning.
First, infeasible candidates are removed, such as those that would trigger out-of-memory (OOM) errors or whose workloads are severely imbalanced relative to the compute capacity of the submesh.
Second, candidates duplicated across repeated modules are marked and pointed to their counterparts (e.g., in Figure~\ref{fig:profiler}(b), a candidate stage in repeated module~2 is pointed to its equivalent in repeated module~1).
After pruning, we apply an off-the-shelf intra-op planner\cite{Alpa, GSPMD} to determine the intra-op parallel strategy for each remaining pair, since intra-op parallelism is confined to homogeneous meshes.
We then profile these candidates to obtain execution information, such as compute and communication costs, and memory footprints, which are stored for subsequent inter-op parallel strategy search.

\subsection{Heterogeneous Inter-op Parallel Strategy Searching Algorithm}
\label{sec:5.2}
We now present an algorithm for searching inter-op parallel strategies to minimize end-to-end training latency on heterogeneous clusters connected by slow interconnects. We first model the end-to-end latency of pipeline training under the H-1F1B scheduler, and then formulate the strategy search as a DP problem to minimize this latency. To make fine-grained layer-level search tractable, we further apply several system-level optimizations, including aggressive pruning and parallelization.

\subsubsection{Execution Order of the Subclusters}
To handle a heterogeneous cluster consisting of $K$ homogeneous subclusters $\{\text{DeviceMesh}_1, \dots, \text{DeviceMesh}_K\}$, we employ a heuristic to determine their sequential execution order. Specifically, for the forward pass, we prioritize placing subclusters with high memory-to-compute ratios at earlier stages. This placement strategy effectively manages the pressure of buffering microbatches, as earlier stages in the 1F1B-style scheduler typically store more intermediate activations.


\subsubsection{Cost Model}
Assume the computational graph is constructed to a layer sequence $(l_1, l_2, \ldots, l_L)$.
We partition this sequence into $S$ stages $(s_1, s_2, \ldots, s_S)$, where each stage $s_i$ is a contiguous subsequence $(l_q, \ldots, l_p)$.
Each stage $s_i$ is then mapped to a submesh $\text{Mesh}_k(n_i, m_i)$ sliced from $\text{DeviceMesh}_k$. Following the rationale in Alpa~\cite{Alpa}, the size of a submesh is constrained to either $(1,1), (1,2), \dots, (1,2^p)$, or $(2, M_k), \dots, (N_k, M_k)$.
Let $t_i = t(s_i, \text{Mesh}_{k}(n_i, m_i))$ denote the compute cost of stage $s_i$ on its assigned submesh, and $c_i$ denote the inter-stage communication cost between stages $i$ and $i{+}1$. Under the H-1F1B scheduler, when $c_i \le t_{\max}$, the end-to-end latency $T^*$ is modeled as:
\begin{align}
T^* &= \min_{\text{strategies}} \left\{ \sum_{i=1}^S (t_i + 2c_i) + (B-1) \cdot \max_{1 \le j \le S} t_j \right\},
\label{eq:gpipe_cost_model}
\end{align}
where the first term captures the cost of a single microbatch (the total time for a forward and backward pass through all $S$ stages). Under the condition $c_i \le t_{\max}$, the H-1F1B scheduling strategy derived in §~\ref{section:4} ensures that the steady phase remains free of communication bubbles. Consequently, the second term accounts for the remaining $(B-1)$ microbatches, which are limited by the slowest stage with latency $t_{\max} = \max_{1 \le j \le S} t_j$. A strategy consists of a sequence of $S$ contiguous stages and their corresponding submesh assignments.

\subsubsection{DP Formulation}
To optimize Eq.~\eqref{eq:gpipe_cost_model}, we design a DP search algorithm. The algorithm enumerates all feasible candidate stage–mesh pairs as a fixed $t_{\max}$ and minimizes the first term in Eq.~\eqref{eq:gpipe_cost_model}, 
Specifically, we define the DP state $F(s, i, d; t_{\max})$ as the minimum execution latency for partitioning the layer subsequence $(l_i, \dots, l_L)$ into $s$ stages using a cumulative total of $d$ GPUs across the ordered subclusters under a given fixed bottleneck constraint $t_{\max}$. To track state of subclusters, we define GPU offsets $\mathcal{O} = [O_0, O_1, \dots, O_K]$, where $O_0 = 0$ and $O_k = \sum_{j=1}^k N_j\cdot M_j$. For a dp state with $d$ total GPUs, the active subcluster $k$ is uniquely identified such that $O_{k-1} < d \le O_k$. The DP transition is:
\begin{align}
\label{eq:dp_general}
&F(s, i, d; t_{\max}) = \\
&\min_{\substack{i \leq j \leq L \\ n\cdot m\le d - O_{k-1}}}
    \left\{
    \begin{array}{l}
         F(s-1, j+1, d - n\cdot m); t_{\max}) + 2C(j)\\
        +\; t((l_i,\dots, l_j), \text{mesh}_{k}(n, m), N \,) 
    \end{array}
    \right\}, \notag
\end{align}
where $C(j)$ denotes the communication cost of transferring intermediate activations from layers $(l_1, \dots, l_j)$ to the subsequent stage. Notably, $C(j)$ depends on both the data size and the interconnect bandwidth between the currently allocated $\text{Mesh}_k$ and the first mesh selected for the following stage. If $d - n \cdot m = O_{k-1}$, the current transition exhausts all remaining GPUs in subcluster $k$, meaning the next stage will reside in $\text{Mesh}_{k-1}$. In this scenario, $C(j)$ is calculated using the inter-cluster bandwidth; otherwise, it is determined by the intra-cluster bandwidth within $\text{Mesh}_k$.

Besides, in Eq.~\eqref{eq:dp_general}, $t((l_i,\dots,l_j), \text{mesh}_k(n,m), N)$ denotes the compute cost of executing the candidate stage $(l_i,\ldots,l_j)$ on $\text{mesh}_k(n,m)$ with N launching counts during warmup. $N$ is a symbolic placeholder of $N(s,i,d;t_{\max})$ and denotes the warm-up launch count of the first stage in this $s$ stage determined by the optimal substructure of $F(s,i,d;t_{\max})$. Specifically, $N$ is used to track the peak GPU memory consumption for each state, ensuring that the generated parallel strategies are feasible and do not trigger Out-of-Memory errors. It can be substituted with the following expression:
\begin{align}
\label{eq:dp_ni}
    N(s,i&,d;t_{\max}) \;=\;\\ &\Big\lceil \tfrac{2\cdot C(j)}{t_{\max}} -\varepsilon \Big\rceil + 1 + N(s-1,j+1,d - n\cdot m; t_{\max}).\notag
\end{align}

For each substructure of $F$, the following constraints must hold:  
(i) $t((l_i,\dots,l_j), \text{mesh}_k(n,m), N) \le t_{\max}$,  
(ii) $C(j) \leq t_{\max}$ (the H-1F1B communication constraint), and  
(iii) the H-1F1B memory constraint as followed: 
\begin{equation}
    \text{mem}_p + N \cdot \text{mem}_a \leq \text{mem}_{\text{device}},
\label{eq:mem_cons}
\end{equation}
where $\text{mem}_p$ denote the memory for model parameters and buffers, $\text{mem}_{\text{device}}$ denote the total device memory, and $\text{mem}_{a}$ denote the activation memory per microbatch.

We initialize
$F(0, L+1, 0; t_{\max}) = 0$ and $N(0, L+1, 0; t_{\max}) = 0$, and the optimal total latency under a fixed $t_{\max}$ is then
\begin{align}
    T(t_{\max}) = 
    \min_{s}\Bigl\{F&\bigl(s,\,0,\,O_K;\,t_{\max}\bigr)\Bigr\} 
    +(B-1)\, t_{\max}. 
\end{align}

\noindent\textbf{Complexity Analysis:}
Assume the heterogeneous cluster consists of $C$ homogeneous DeviceMeshes.  
Ignoring variations in DeviceMesh sizes, the slicing step of our DP algorithm under a fixed $t_{\max}$ requires
$O\bigl(CL^3 NM(N + \log M)\bigr)$
time. 
The number of candidate $t_{\max}$ values produced by the profiler is upper-bounded by
$O\bigl(CL(N + \log M)\bigr)$. 
Since the profiling across different homogeneous DeviceMeshes can be parallelized, the effective complexity of this step is reduced to  
$O\!\left(L(N + \log M)\right)$.  
Thus, the overall complexity of the search algorithm is
$O\bigl(CL^4 NM(N+\log M)^2)$.

\subsubsection{Search Overhead Optimizations}
Running the DP search at fine-grained layer granularity (e.g., $L \!\sim\! 10^2$) leads to a combinatorial explosion of the state space, requiring hundreds of hours to find an optimal strategy. To make the search tractable, we introduce three optimizations.  
\textbf{(1) DP sparsity index.} The theoretical state space of $F$ scales as $O(L^2N^CM^C)$. However, the Zero-Redundant Profiler prunes a large fraction of infeasible candidate stage-mesh pairs in advance. We exploit this by pre-computing a sparse index of feasible candidate stage-mesh pairs and restricting DP transitions to this reduced set, thereby avoiding redundant states.  
\textbf{(2) Bidirectional pruning of $t_{\max}$.} Candidate $t_{\max}$ values are first sorted and deduplicated from profiler outputs. We then locate the smallest feasible value $t_S$ via binary search and discard all smaller ones. From the solution at $t_S$ with latency $T(t_S)$, we derive an upper bound $t_E=\lceil T(t_S)/B \rceil$. Any $t_{\max} > t_E$ is pruned, as larger values cannot further reduce latency.  
\textbf{(3) Batched parallel evaluation.} The remaining $t_{\max}$ candidates are evaluated in parallel using Ray actors. Instead of launching one task per candidate—which causes imbalance and Just-in-time overhead—we batch candidates by grouping them according to the number of valid stage–mesh pairs activated. This balances the per-actor workload and amortizes warm-up costs.

%% file: sections/06-evaluation.tex
\section{Evaluation}
\label{section:6}
\begin{figure*}[t]
    \centering
    \includegraphics[width=\linewidth]{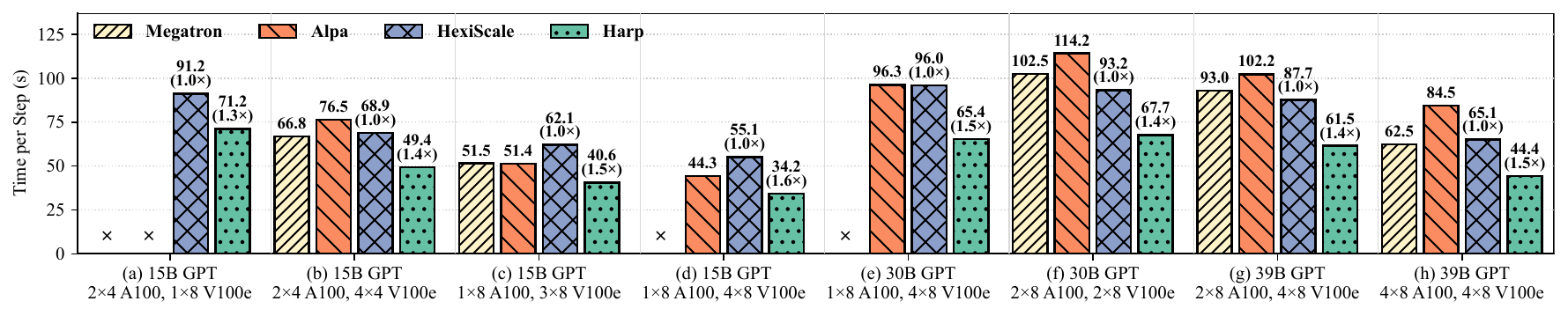}
    \caption{End-to-end training latency of {Harp} compared with baselines on GPT model.}
    \label{fig:evaluation_e2e_heter}
\end{figure*}

\begin{figure}[t]
    \centering
    \includegraphics[width=\linewidth]{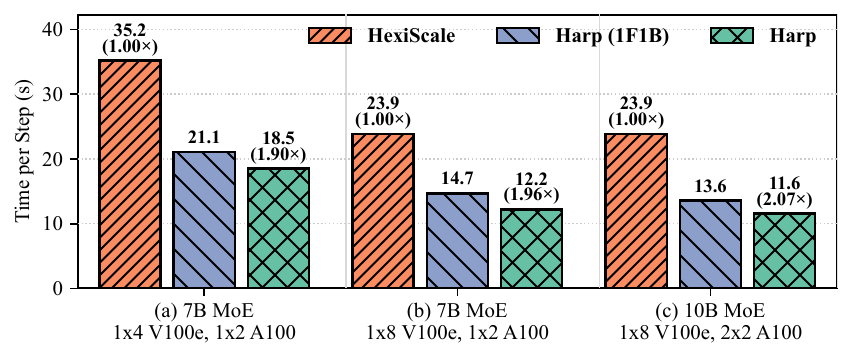}
    \caption{End-to-end training latency of {Harp} compared with baselines on MoE model.}
    \label{fig:evaluation_moe}
\end{figure}



We implement \textsc{Harp} on top of Alpa, extending it with 3{,}000+ lines of Python and C++ code.
Beyond extending Alpa’s compiler with our planner and pipeline scheduler, 
we implement a heterogeneity-aware runtime on top of Ray actors~\cite{moritz2018ray}, XLA~\cite{XLA}, and NCCL~\cite{NCCL}, 
which adapts to diverse interconnects by automatically selecting the fastest available Network Interface Card (NIC) based on the network topology.

Our evaluation is conducted on 3 heterogeneous settings:

\noindent\textbf{Heterogeneous Setting 1:} A cluster comprising 64 GPUs, consisting of 4 nodes each equipped with 8$\times$NVIDIA A100 GPUs and 4 nodes each equipped with 8$\times$NVIDIA V100e GPUs\footnote{To emulate V100e performance on A100 nodes, we throttle the GPU core frequency and cap the maximum memory capacity.}. Detailed hardware specifications are summarized in Table~\ref{tab:gpu_specs_transposed}.

\noindent\textbf{Heterogeneous Setting 2:} A cluster with 12 GPUs across two distinct configurations: 1 node with 8$\times$V100 (16GB, NVLink), 2 node with 2$\times$A100 (40GB, PCIe).

\noindent\textbf{Heterogeneous Setting 3:} A cluster with 12 GPUs across three distinct configurations: 1 node with 8$\times$V100 (16GB, NVLink), 1 node with 2$\times$V100e (32GB, PCIe), and 1 node with 2$\times$A100 (40GB, PCIe).

For both settings, each GPU is allocated 8 CPU cores with 8\,GB of RAM per core. Intra-cluster communication within each homogeneous subcluster is performed via RDMA over Mellanox ConnectX-6 NICs, while inter-cluster communication is conducted over a 10\,Gbps Ethernet network.

\noindent\textbf{Models.}  
We evaluate \textsc{Harp} on GPT models ranging from 15B to 39B parameters and MoE models ranging from 7B to 10B scaling the hidden dimension with the number of available devices. The global batch size is fixed at 1024 and the sequence length at 1K.

\subsection{Comparison under Heterogeneous Clusters}

\subsubsection{End-to-End Performance} \mbox{}\\
\label{section:6.1}

\noindent\textbf{GPT Model.} 
We evaluate end-to-end training latency under eight heterogeneous configurations on setting 1 and three GPT model sizes (Figure~\ref{fig:evaluation_e2e_heter}), comparing \textsc{Harp} against three representative systems.  
All baselines are tuned with their best hyperparameter settings, and experiments are conducted with cross-cluster bandwidth fixed at 5\,Gbps\footnote{Lower-bandwidth links are emulated by using Linux trafic control (\texttt{tc}).}. 
Note that Megatron and HexiScale leverage FlashAttention optimizations~\cite{dao2022flashattention, dao2023flashattention2}, whereas Alpa and \textsc{Harp} rely solely on JAX’s JIT compilation~\cite{Jax}. As a result, \textsc{Harp} is at a computational disadvantage in attention-intensive blocks. These attention optimizations, however, are orthogonal to the framework-level techniques studied in this work.

As shown in Figure~\ref{fig:evaluation_e2e_heter}(a), neither Alpa nor Megatron supports configurations where the number of devices differs across nodes, as both are designed for homogeneous clusters. In Figure~\ref{fig:evaluation_e2e_heter}(d) and (e), \emph{Megatron} fails to identify a valid parallel strategy that fully utilizes all available devices due to its limited parallel strategy search space.
Overall, despite lacking kernel-level optimizations, \textsc{Harp} consistently outperforms the best baseline (HexiScale) by 1.3×–1.6× across all evaluated heterogeneous configurations, demonstrating its effectiveness in heterogeneous environments.

\noindent\textbf{MoE Model.} We evaluate the end-to-end performance of MoE models under Heterogeneous Setting 2. In addition to comparing with HexiScale, we conduct an ablation study by replacing our H-1F1B scheduler with a native 1F1B scheduler within \textsc{Harp}. As shown in Figure~\ref{fig:evaluation_moe}, \textsc{Harp} with H-1F1B achieves a $1.14\times$--$1.20\times$ speedup over the native 1F1B configuration. This improvement stems from the fact that \textsc{Harp}’s joint optimization of parallel strategy and pipeline scheduling explicitly accounts for both workload balancing across heterogeneous devices and communication-compute overlap across heterogeneous networks. For instance, in the 10B MoE experiment (8$\times$V100e + 4$\times$A100), \textsc{Harp} partitions the model into four pipeline stages and generates a non-uniform warm-up sequence of $[5, 4, 2, 1]$ forward microbatches. This schedule ensures fully communication hiding during the steady phase, significantly enhancing e2e throughput.

Furthermore, even when both systems use the 1F1B scheduler, \textsc{Harp}’s planner outperforms HexiScale by $1.63\times$--$1.75\times$ on MoE workloads—a more pronounced advantage than observed in GPT models. This is primarily due to HexiScale’s limited inter-op planning layer granularity. HexiScale treats a standard decoder layer as the atomic unit for inter-op planning and assumes all layers are identical. However, the MoE models in our evaluation consist of interleaved MoE and standard decoder layers. To accommodate this, HexiScale is forced to logically group these distinct layers into pairs, reducing the effective search space to only 8 units for an 8+8 layer model. In contrast, \textsc{Harp}’s planner can handle heterogeneous layer types and partitions the model at a much finer granularity (e.g., 50 layers). This expanded search space allows \textsc{Harp} to more effectively balance computational tasks with the diverse compute capacities of heterogeneous device.

\subsubsection{Other Baseline.}

We do not include Metis~\cite{um2024metis} in our comparative analysis primarily due to its architectural constraints. First, the current implementation of the Metis planner requires a uniform number of GPUs across all nodes, including heterogeneous ones. This constraint is incompatible with many real-world heterogeneous scenarios, such as our Heterogeneous Setting 2 and 3, where enforcing such uniformity would restrict the usable GPU count to only 6 instead of the full 12, leading to significant throughput degradation. Second, prior evaluations by Yan et al.\cite{yan2024hexiscale} have demonstrated that HexiScale consistently outperforms Metis, achieving performance improvements of $1.6\times$ to $1.9\times$. Given that HexiScale already represents a state-of-the-art advancement over Metis, we focus our comparative analysis on HexiScale.

\subsubsection{Pipeline Breakdown Analysis.}  
\begin{figure}[t]
    \centering
    \includegraphics[width=\linewidth]{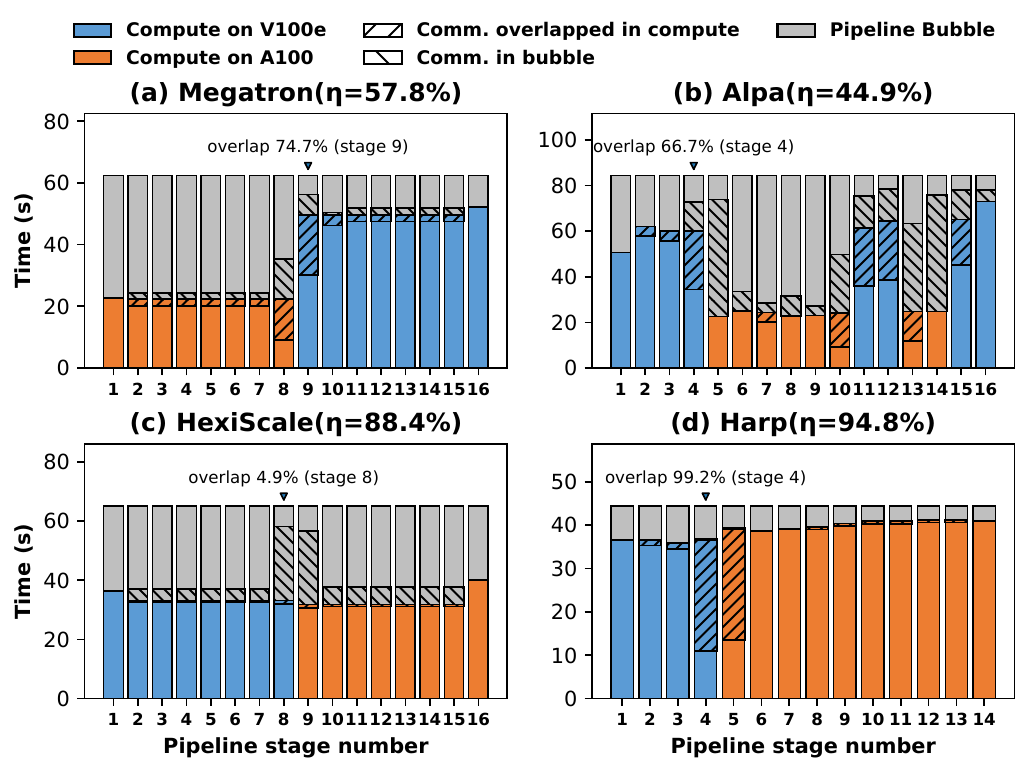}
    \caption{Stage-wise breakdown of computation, communication, and idle bubbles within one iteration for the configuration in Figure~\ref{fig:evaluation_e2e_heter}(h). 
    For each framework, we annotate the communication–latency overlap ratio at the stage located on the subcluster boundary.}
        
    \label{fig:evaluation_pipeline_breakdown}
\end{figure}

Figure~\ref{fig:evaluation_pipeline_breakdown} shows the stage-wise breakdown of computation, communication, and idle time (pipeline bubbles) under the same configuration as Figure~\ref{fig:evaluation_e2e_heter}(h). Communication latency reflects only active data transfers on NICs, while idle waiting is attributed to bubbles. This breakdown analysis reveals the sources of \textsc{Harp}’s performance advantage.

\noindent\textbf{Load Balance Analysis.}  
To quantify the degree of imbalance, we define the load-balancing metric $\eta$: 
\begin{equation}
\eta = 1 - \frac{\sum_{i=1}^d (td_{\max} - td_i)\cdot \text{PeakFlops}_i}{td_{\max} \cdot \sum_{i=1}^d \text{PeakFlops}_i} \times 100\% ,
\end{equation}
where $d$ is the number of devices, $td_i$ is the actual compute time of device $i$, $\text{PeakFlops}_i$ is its theoretical peak FLOP/s, and $td_{\max} = \max_{i=1}^{d} td_i$ is the maximum compute time among all devices. $\eta$ ranges from $(0\%, 100\%]$, with $\eta = 100\%$ indicating perfect balance, i.e., identical compute times across all devices. We adopt $\eta$ in our evaluation to measure how well different strategies balance heterogeneous workloads.

From the results, both Megatron and Alpa overlook the performance gap between A100 and V100e GPUs. For example, in Figure~\ref{fig:evaluation_pipeline_breakdown}(a), stages mapped to V100es in Megatron take up to $2.6\times$ longer than those on A100s. In contrast, in Figure~\ref{fig:evaluation_pipeline_breakdown}(c) and (d), HexiScale and \textsc{Harp} allocate workloads proportionally to device capabilities, thereby reducing idle time. However, HexiScale performs inter-op planning at coarse layer granularity; in this example, it aggregates the embedding and output layers into boundary stages, thereby inflating their workload and reducing load-balancing score $\eta$ to $88.4\%$. By comparison, \textsc{Harp} deliberately assigns fewer layers to slower V100es to improve the throughput of faster A100s, striking a better balance with  $\eta=94.8\%$ and thereby achieving higher end-to-end performance.

\noindent\textbf{Overlap Analysis.}  
Beyond load balancing, differences in communication–computation overlap also significantly impact performance.  
As shown in Figure~\ref{fig:evaluation_pipeline_breakdown}(a), \emph{Megatron} overlaps 74.7\% of communication latency. This is because, under data-parallel training, where multiple \texttt{processGroup}s manage different pipelines, inter-stage communication from one pipeline can overlap with computation from another.
In contrast, Figure~\ref{fig:evaluation_pipeline_breakdown}(c) shows that \emph{HexiScale} overlaps only 4.9\%. To support heterogeneous pipeline parallelism, it enforces strict synchronization across pipelines, thereby losing the opportunity for overlap provided by data parallelism. The small amount of overlap observed is implicitly contributed by TCP buffer effects.  
Figure~\ref{fig:evaluation_pipeline_breakdown}(b) illustrates that \emph{Alpa}, which uses SPMD-style data parallelism, cannot exploit data parallelism for overlap either. However, by employing Eager-1F1B scheduling, it achieves 66.7\% overlap.  
Finally, as shown in Figure~\ref{fig:evaluation_pipeline_breakdown}(d), \textsc{Harp} leverages the adaptive H-1F1B scheduler. In this case, since $c < t_{\max}$, almost all communication is overlapped. Notably, unlike other frameworks, \emph{Alpa} uses Ethernet NICs for communication across homogeneous subclusters but RDMA within each subcluster; this benefit comes from its heterogeneous runtime, which adaptively selects the appropriate network interface.

\subsection{Performance Comparison on Heterogeneous and Homogeneous Clusters}
\begin{figure}[t]
    \centering
    \includegraphics[width=\linewidth]{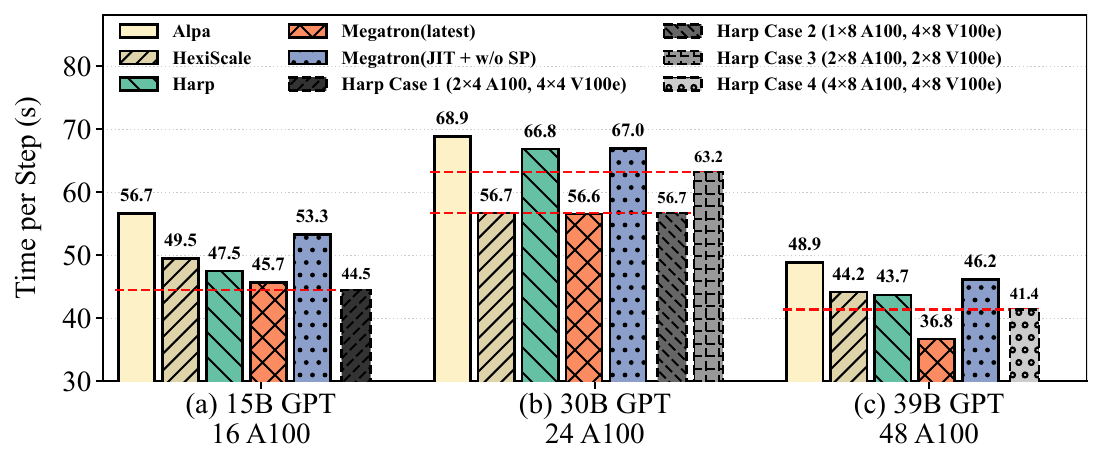}
    \caption{Comparison of end-to-end training latency across homogeneous and heterogeneous GPU clusters.}
    \label{fig:evaluation_hetero_vs_homo}
\end{figure}

We evaluate the end-to-end training latency of Megatron, Alpa, HexiScale, and Harp on homogeneous clusters, and use these results as the baseline. We then compare \textsc{Harp}’s performance on heterogeneous clusters (Figure~\ref{fig:evaluation_e2e_heter}(b,e,f,h)) against this homogeneous baseline to assess its ability to fully exploit heterogeneous resources.
Each heterogeneous cluster is chosen such that its theoretical peak FLOP/s lies within 87–93\% of that of the corresponding homogeneous cluster. Latency results in Figure~\ref{fig:evaluation_hetero_vs_homo} are normalized by this ratio to ensure fairness.
For completeness, we further distinguish two versions of Megatron: (i) the latest fully optimized release, and (ii) a conservative lower-bound configuration using only PyTorch JIT with sequence parallelism disabled.

As shown in Figure~\ref{fig:evaluation_hetero_vs_homo}(c), we compare \textsc{Harp} on a heterogeneous cluster with its homogeneous baseline. 
Although the heterogeneous cluster provides only 93\% of the peak FLOP/s of the homogeneous one, \textsc{Harp} still sustains 82.7\% of Megatron’s throughput.
In an even more extreme configuration (Figure~\ref{fig:evaluation_hetero_vs_homo}(a)), \textsc{Harp} on a heterogeneous cluster even outperforms all homogeneous baselines. This advantage arises from the larger aggregate memory of the heterogeneous cluster, which allows for higher degrees of data parallelism. These results highlight that \textsc{Harp} can fully exploit heterogeneous compute resources, even in the presence of slow network bottlenecks.


\subsection{Scalability across Subcluster Number}
\begin{figure}[t]
    \centering
    \includegraphics[width=\linewidth]{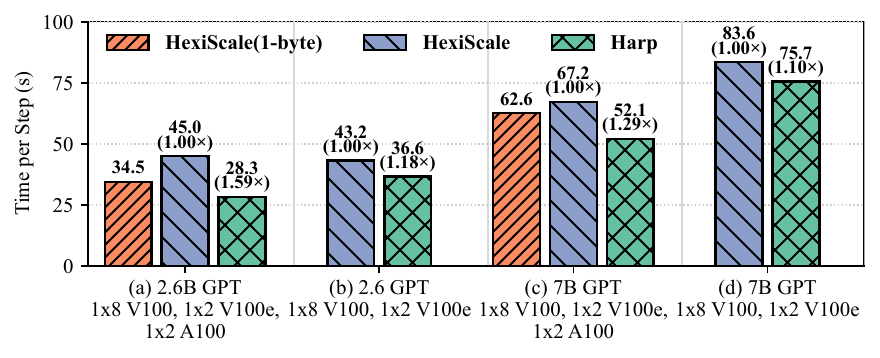}
    \caption{Comparison of end-to-end training latency for heterogenous cluster with 3 up to homogeneous sub-cluster.}
    \label{fig:evaluation_muti}
\end{figure}

To assess how \textsc{Harp} handles increasing heterogeneity and cluster scale, we compare its performance with HexiScale using Heterogeneous Setting 3. Specifically, we evaluate two configurations: a dual-subcluster setup (V100 + V100e) and a triple-subcluster setup (V100 + V100e + A100), using GPT-2.6B and GPT-6.7B as workloads. In the triple-subcluster setup, the A100 nodes are connected to the V100/V100e nodes via 5\,Gbps links, whereas the V100 and V100e subclusters are interconnected via a high-speed 200\,Gbps IB network. 
Additionally, we introduce an alternative baseline for HexiScale in the triple-subcluster configuration by reducing all inter-stage data transfers to a 1-byte tensor. This baseline eliminates pipeline communication overhead, allowing us to isolate and evaluate the intrinsic quality of the parallel strategies generated by \textsc{Harp}’s and HexiScale’s planners, independent of pipeline communication overhead.

As shown in Figure~\ref{fig:evaluation_muti}, \textsc{Harp} achieves a $1.1\times$--$1.2\times$ speedup over HexiScale in both the dual-subcluster setup and the 1-byte communication control case. Furthermore, we profile the execution and find that the performance gap widens significantly in highly heterogeneous scenarios. This divergence stems from \textsc{Harp}’s ability to perform fine-grained inter-op partitioning, which enables task allocations to precisely match the heterogeneous compute capacities of each submesh. These results demonstrate that by searching at a finer granularity, \textsc{Harp}’s strategy planner can identify more balanced parallel strategies that remain unattainable for existing coarse-grained methods in complex environments.

Furthermore, in the V100+V100e+A100 setting, \textsc{Harp}'s runtime dynamically selects network interface cards (NICs) based on the specific interconnect environment and adaptively tunes its pipeline scheduling strategy. In contrast, HexiScale relies on the native 1F1B scheduler, which fails to overlap stage-to-stage communication overhead. Moreover, HexiScale’s network selection is constrained: it must uniformly employ either IB or Ethernet across all nodes. If even a single node lacks an IB connection, the high-speed IB interfaces on all other nodes remain underutilized. Consequently, \textsc{Harp} outperforms HexiScale by $1.3\times$--$1.6\times$ in end-to-end training throughput.

\subsection{Sensitivity to Cross-Cluster Bandwidths}
\begin{figure}[t]
    \centering
    \includegraphics[width=\linewidth]{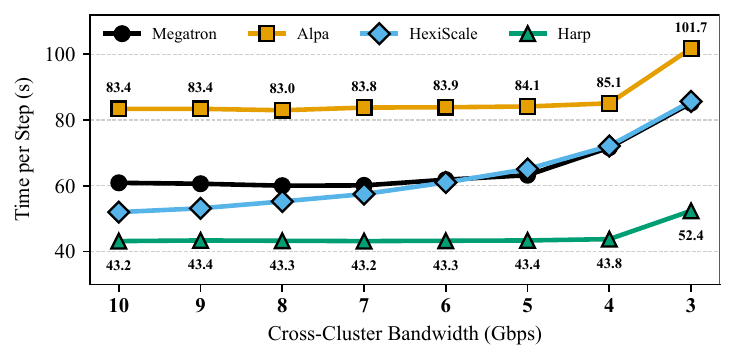}
    \caption{
    End-to-end training latency of different systems under varying cross-cluster bandwidths (3–10\,Gbps). 
    }
    \label{fig:evaluation_network_sweep}
\end{figure}

To evaluate the impact of cross-cluster bandwidth, we use the heterogeneous configuration in Figure~\ref{fig:evaluation_e2e_heter}(h) and throttle the cross-cluster links from 10\,Gbps down to 3\,Gbps using Linux \texttt{tc}. 
As shown in Figure~\ref{fig:evaluation_network_sweep}, HexiScale is highly sensitive to bandwidth, with latency growing roughly inversely proportional to the available bandwidth, since it cannot explicitly overlap communication with computation. 
Megatron relies only on data parallelism and TCP buffering to implicitly overlap communication. As a result, when bandwidth falls below 7\,Gbps, its trend converges with that of HexiScale.  
In contrast, \textsc{Harp} maintains stable step time until the bandwidth drops to around 3\,Gbps, at which point $c > t_{\max}$ and the inter-stage communication overhead surpasses the total computation cost. 
Interestingly, Alpa exhibits a similar inflection point as \textsc{Harp}, suggesting that under this configuration, Eager-1F1B and H-1F1B have comparable sensitivity to cross-cluster bandwidth. However, this does not imply equivalent overlap capability. As shown in Figure~\ref{fig:evaluation_pipeline_breakdown}, Alpa suffers from substantial load imbalance that doubles $t_{\max}$ compared to \textsc{Harp}, which masks the limitations of its weaker overlap strategy in this experiment.

\subsection{Ablation Study of Layer Granularity}
\label{section:6.3}

We examine the effectiveness of applying heterogeneous inter-op parallel strategy planning at different layer granularities. 
Using the heterogeneous configuration in Figure~\ref{fig:evaluation_e2e_heter}(h), we evaluate three baseline granularities with \textsc{Harp}’s runtime including \#layer-8, \#layer-16 and \#layer-48. 
As shown in Figure~\ref{fig:evaluation_ablation}(a), strategies planned on \textsc{Harp}’s fine-grained layer sequence consistently outperform all baselines by 1.2$\times$–1.6$\times$, demonstrating the necessity of fine-grained decomposition.  


\subsection{Ablation Study of Joint Optimization}
\begin{figure}[t]
    \centering
    \includegraphics[width=\linewidth]{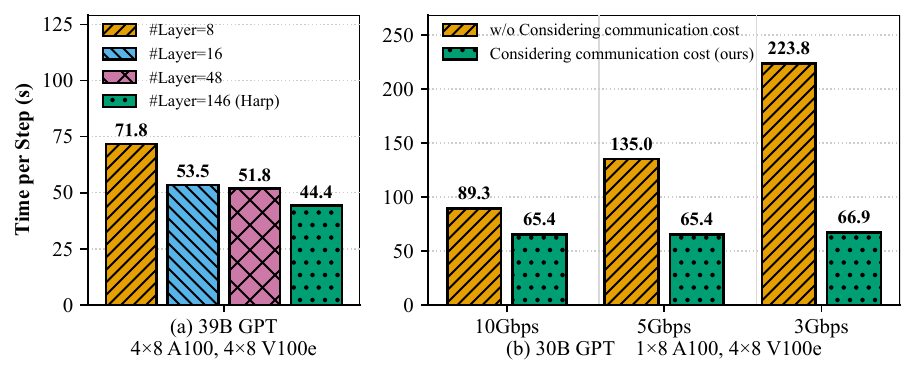}
    \caption{(a) Effect of fine layer granularity, (b) Effect of joint optimization of planning and scheduling.}    \label{fig:evaluation_ablation}
\end{figure}

We evaluate the effectiveness of jointly optimizing parallel strategy planning and pipeline scheduling under the heterogeneous configuration in Figure~\ref{fig:evaluation_e2e_heter}(e). 
Specifically, we compare \textsc{Harp} with a baseline that \emph{ignores communication cost}, where parallel strategies are planned assuming zero communication overhead, i.e., by setting $C(i) = 0$ for all $i \in \{1, \ldots, s\}$. As shown in Figure~\ref{fig:evaluation_ablation}(b), the performance of baseline is substantially worse than \textsc{Harp}, resulting in a 1.4$\times$–3.3$\times$ slowdown. 
This is because the size of tensor transferred across stages can vary substantially depending on the chosen inter-op parallel strategy. 
We inspected the inter-op parallel strategy generated by this baseline, and observed that it achieved a high load-balancing score ($\eta = 94.8\%$). 
However, it assigned two large tensors to the bottleneck link, incurring $5\times$ higher communication overhead than our method.
In contrast, \textsc{Harp}’s strategy transmits only the smaller $(\text{Batch\_Size}, \text{Seq\_Len}, \text{Hid\_Dim})$ tensor over the bottleneck. 
Thus, even with a lower load-balancing score ($\eta = 93.3\%$), \textsc{Harp} delivers markedly better end-to-end performance. 
These results demonstrate that explicitly accounting for communication cost is essential for planning on heterogeneous networks.

\subsection{Search Overhead}
\label{search overhead}
We evaluate the planning overhead of a heterogeneous configuration in Figure~\ref{fig:evaluation_e2e_heter} (h). 
Without our optimizations (Alpa’s default mechanisms), profiling alone takes more than 10 hours and the DP search exceeds 100 hours, making planning impractical. 
In contrast, \textsc{Harp} completes profiling in 1248 seconds and finishes the DP search in just 133 seconds, reducing the total overhead to about 23 minutes. 
We further evaluate the scalability of our planner under heterogeneous configurations, as shown in Figure~\ref{fig:evaluation_e2e_heter}(e–h). As the model and cluster scale up, the planning overhead actually decreases (47 min $\rightarrow$ 44 min $\rightarrow$ 33 min $\rightarrow$ 23 min). This trend arises because both GPT-30B and GPT-39B share the same layer granularity of $\#Layer = 146$. While larger clusters introduce additional profiling and searching complexity, they also increase the proportion of candidate stage–mesh pairs that can be pruned. In addition, the number of CPU cores available for parallel profiling and searching grows with cluster size, further reducing the overhead.
These results demonstrate that our layer-construction method, zero-redundant profiler, and system-level optimizations for dynamic programming search make fine-grained heterogeneous planning practical and efficient for large-scale model training.


%% file: sections/07-discussion.tex
\section{Discussion}


\noindent\textbf{Fault Tolerance and Dynamic Preemption.} \textsc{Harp} can rapidly regenerate heterogeneous parallel strategies when the underlying training infrastructure changes, as demonstrated by our evaluation of search overhead in §\ref{search overhead}. However, the current implementation does not yet support recovery from hardware preemption or failures during runtime. Fully addressing such scenarios would require an integrated framework for automatic checkpointing and dynamic re-profiling. We consider the development of a fault-tolerant, elastic scheduling system for heterogeneous environments as an important direction for our future work.


\noindent\textbf{Impact of Resource Contention.} In frameworks that support varied data and tensor parallelism across stages, collective communication (e.g., broadcast) is often preferred for inter-stage transfers \cite{Alpa-opt}. However, such primitives consume more GPU resources than P2P kernels, often degrading the performance of concurrent computations. While we currently mitigate this interference by capping {NCCL\_MAX\_CTAS}, this static approach is suboptimal for heterogeneous clusters. Different GPU architectures exhibit vastly different SM counts and scheduling behaviors, necessitating a more dynamic, architecture-aware resource partitioning strategy to balance communication and compute efficiency.

%% file: sections/08-related.tex
\section{Related Work}

Many recent researches have explored optimized parallel training strategies tailored for heterogeneous clusters. HAP~\cite{zhang2024hap} extends the SPMD (Single Program Multiple Data) paradigm to heterogeneous accelerators by introducing adaptive sharding ratios for tensor dimensions, while Hetu-V2~\cite{li2025hetu} proposes a unified H-SPMD abstraction to manage heterogeneous workloads across various hardware types. To address heterogeneity on inter-op level, PipePar~\cite{zhang2023pipepar}, AMP~\cite{li2022amp}, and HetHub~\cite{xu2024hethub} partition computation graphs into pipeline stages with uneven workload, ensuring balanced execution across accelerators with varying capabilities. Metis~\cite{um2024metis} incorporates heterogeneity into both data and pipeline parallelism dimensions. Expanding on this, HexiScale~\cite{yan2024hexiscale} further introduces asymmetric tensor partitioning and allows for stage-wise variations in both the number of decoder layers and the degree of tensor parallelism. 
More recently, a concurrent work, Sailor~\cite{strati2025sailor}, introduced an automated distributed training framework tailored for dynamic, heterogeneous, and geo-distributed clusters. Sailor specializes in system elasticity, employing a sub-second strategy search to enable rapid, automated fault recovery and adaptation within volatile resource pools.

For pipeline schedulers, early systems like GPipe~\cite{Gpipe} and PipeDream~\cite{PipeDream, PipeDream-2BW} extended native model parallelism into pipelining. Megatron\cite{Megatron-lm-v2} proposed interleaved pipelines, while Chimera~\cite{li2021chimera} introduced bidirectional pipelines, both reducing pipeline bubbles. Eager-1F1B~\cite{Alpa-opt} increases warm-up launching counts to overlap computation and communication in the steady phase, and Zero-Bubble pipeline~\cite{qi2023zero} further splits backward passes for finer-grained overlap.

%% file: sections/09-conclusion.tex
\section{Conclusion}

We present \textsc{Harp}, an automated parallel training framework for heterogeneous clusters. 
\textsc{Harp} introduces an efficient fine-grained parallel strategy planner that mitigates communication overhead while maintaining balanced loads across heterogeneous accelerators. 
With a heterogeneity-aware pipeline scheduler, Harp achieves the theoretical upper bound of computation–communication overlap with minimal memory overhead.
Our evaluation demonstrates that \textsc{Harp} delivers $1.3\times$–$1.6\times$ higher performance than state-of-the-art frameworks designed for model training on heterogeneous clusters.

\section{Acknowledgements}
We thank the anonymous reviewers and our shepherd for their insightful feedback. We also thank the National Supercomputing Center (Jinan) team for their technical support. This work was supported by the Project of Key R\&D Program of Shandong Province (Grant No. 2024CXGC010113). We are also grateful to Jingfan Yao and Tian Jiang for providing essential network service support. Furthermore, we appreciate the helpful discussions and feedback from Foteini Strati, Fangcheng Fu, Ran Yan, Lixia Chen, Yangguang Shao, and Hangyu Zhang.

%% file: EuroSys26_ArtifactAppendix_template.tex
%

\newpage

\appendix
\section{Artifact Appendix}

\subsection{Abstract}
The HARP artifact includes the source code of the heterogeneity-aware planner, the adaptive H-1F1B runtime scheduler, and the benchmarking suite used in the paper. It is built upon the Alpa, JAX, and Ray. We provide a Docker-based environment to simplify the setup of heterogeneous GPU clusters and include scripts to reproduce the automated parallelization planning and scheduling logic.

\subsection{Description \& Requirements}

\subsubsection{How to access}
This artifact is archived on Zenodo (\url{https://doi.org/10.5281/zenodo.18909673}) and Github (\url{https://github.com/Lssyes/harp})

\subsubsection{Hardware dependencies}
A minimum of two nodes is required, each equipped with at least one NVIDIA GPU. HARP is designed for heterogeneous setups (e.g., A100 + V100 nodes or RTX 4080 + 4090 nodes). The nodes must be interconnected via TCP/IP.

\subsubsection{Software dependencies}\
The artifact relies on Docker (20.10+) and the NVIDIA Container Toolkit. All other dependencies, including CUDA 11.3/11.8, JAX, and Ray, are encapsulated within the provided Dockerfile.

\subsubsection{Benchmarks} 
The artifact uses GPT-3 model configurations of various sizes as the primary workload. 
Pre-configured gin files are provided in the \path{harp_patches/alpa/config/} directory. 
Additionally, we provide a wide range of search parameters for different heterogeneous cluster configurations in \path{benchmark/suite_auto_gpt.py}. 
Users can switch between different suites by modifying the arguments of \path{benchmark.py}.

\subsection{Set-up}

The set-up process consists of environment preparation, Docker image building, and Ray cluster initialization. We provide a Docker-based encapsulation to ensure environment consistency across heterogeneous nodes.

\noindent\textbf{Environment Preparation}

\noindent\textbf{1. Clone the Repository:} Clone the source code along with all submodules:

\begin{verbatim}
git clone --recursive  \
https://github.com/Lssyes/harp.git && \
cd harp
\end{verbatim}

\noindent\textbf{2. SSH Key Generation:} HARP requires SSH for inter-node coordination. Generate a key pair in the specific directory before building the image:
\begin{verbatim}
mkdir -p docker/ssh_key
ssh-keygen -t rsa -b 4096 \
-f ./docker/ssh_key/id_rsa -N ""
cat ./docker/ssh_key/id_rsa.pub >> \
./docker/ssh_key/id_rsa/authorized_keys
\end{verbatim}

\noindent\textbf{Building Docker Images}
On each node, build the Docker image matching its specific GPU architecture (e.g., SM 70 for V100, SM 80 for A100). Refer to the README table for the correct \texttt{BASE\_IMAGE} and \texttt{SM} arguments:
\begin{verbatim}
docker build --build-arg BASE_IMAGE="$BASE_IMAGE" \
             --build-arg SM="$SM" \
             -f docker/build_harp.Dockerfile \
             -t "harp:sm$SM" .
\end{verbatim}

\noindent\textbf{Profiling and Cluster Launch}

\noindent\textbf{1. Container Startup:} Launch the container on all nodes. If a shared NFS is available, mount it to \texttt{/workspace/nfs\_share} to facilitate profile database sharing.
\begin{verbatim}
docker run -it --rm --network host --shm-size 24G \
    --privileged --gpus all \
    -v /path/to/nfs:/workspace/nfs_share harp:sm$SM
\end{verbatim}

\noindent\textbf{2. Hardware Profiling:} Run \texttt{gen\_prof\_database.py} on one representative node for each GPU type to generate the cost model (\texttt{.pkl} files). Ensure these files are accessible to all nodes via NFS or manual \texttt{scp}.

\noindent\textbf{3. Ray Initialization:} On the head node (\texttt{ip\_list[0]}), edit \path{scripts/ray_cluster.sh} to include the cluster IP addresses and GPU counts, then execute:
\begin{verbatim}
bash /workspace/harp/scripts/ray_cluster.sh
\end{verbatim}
Select option `1' when prompted to start the heterogeneous cluster.

\subsection{Evaluation workflow}

\noindent\textbf{Major Claims}
\begin{itemize}
    \item {(C1): HARP can automatically generate efficient parallelization strategies for heterogeneous clusters (A100 + V100). Proven by End-to-End Performance (figure7).}
    \item {(C2): The H-1F1B scheduler achieve better compute-communicaiton overlaping across stages on slow network. Proven by experiment Pipeline Breakdown Analysis(figure 8) and exportment Sensitivity to Cross-Cluster Bandwidths(figure 10).}
\end{itemize}

\noindent\textbf{Experiments}

\noindent\textbf{Experiment (E1): End-to-End Performance Verification.} 
This experiment compares the end-to-end training latency of HARP across various heterogeneous settings against baseline frameworks, corresponding to Figure 7 in the paper.

\begin{itemize}
    \item \textbf{[Preparation]} Ensure the Ray cluster is initialized and profiling databases (\texttt{.pkl} files) for the target heterogeneous GPUs (e.g., A100 and V100) are placed in the shared directory.
    \item \textbf{[Execution]} On the head node, execute the search and training script with the heterogeneity-aware planner enabled(refered in README)
    \item \textbf{[Results]} The system will output an optimized execution plan (sharding and pipeline stages) and a tailored pipeline scheduling strategy.

\end{itemize}

\noindent\textbf{Experiment (E2): Pipeline Breakdown and Overlap Analysis.} 
This experiment uses Nsight Systems to profile the execution and overlap of computation and communication streams under different scheduling policies, corresponding to Figure 8.

\begin{itemize}
    \item \textbf{[Execution]} Modify the \texttt{gin-file} and \texttt{harp/alpa/global\_config.py} to toggle scheduling policies. Three configurations should be tested:
        \begin{enumerate}
            \item \textbf{H-1F1B (HARP):} Set \texttt{enable\_H1F1B = True} and \texttt{global\_config.enable\_overlapping = True}.
            \item \textbf{Overlap-friendly (Alpa):} Set \texttt{enable\_H1F1B = False} and \texttt{global\_config.enable\_overlapping = True}.
            \item \textbf{Vanilla 1F1B:} Set both to \texttt{False}.
        \end{enumerate}
    Run the benchmark with \texttt{nsys profile} to capture the timeline.
    \item \textbf{[Results]} By inspecting the Nsight Systems UI, researchers should observe that H-1F1B achieves superior overlap between computation and cross-mesh communication compared to the baselines, as illustrated in the breakdown analysis of Figure 8.
\end{itemize}

\noindent\textbf{Experiment (E3): Sensitivity to Cross-Cluster Bandwidth.} 
This experiment evaluates the robustness and performance gains of H-1F1B under varying network bandwidths between heterogeneous sub-clusters, corresponding to the sensitivity analysis in Figure 9.

\begin{itemize}
    \item \textbf{[Preparation]} Use the provided bandwidth control scripts (located in \path{/workspace/scripts/network_sim.sh}) to simulate different cross-cluster interconnect speeds (e.g., 3-10Gbps).
    \item \textbf{[Execution]} Repeat the execution steps from Experiment (E2) for each simulated bandwidth setting.
    \item \textbf{[Results]} The resulting end-to-end performance should demonstrate that HARP’s H-1F1B scheduler provides higher speedups as the asymmetry between inter-node and intra-node bandwidth increases, matching the trends in the paper.
\end{itemize}


%% file: appendix/A-proof.tex
\section{Extend proof of S-stage Balanced Pipeline Case}
\label{Appendix:A}
In this appendix, we provide the detailed derivation of the steady-phase latency for an $S$-stage balanced pipeline and prove that the H-1F1B scheduling strategy minimizes this latency.
\begin{table*}[ht]
    \centering
    \caption{MFU of Figure-\ref{fig:evaluation_e2e_heter}.}
    \label{tab:mfu}
    \begin{tabular}{lcccccccc}
        \toprule
        \textbf{Heterogeneous Setting} & \textbf{(a)} & \textbf{(b)} & \textbf{(c)} & \textbf{(d)} & \textbf{(e)} & \textbf{(f)} & \textbf{(g)} & \textbf{(h)} \\
        \midrule
        {Megatron}  & N/A & 44.1\%	&46.8\%& N/A& N/A	&35.8\% & 39.9\%&	38.2\% \\
        {Alpa}      & N/A & 38.5\% &46.9\% &46.0\% &41.0\% &32.1\% &36.3\% &28.2\% \\
        {HexiScale} &41.5\%	&42.8\%	&38.8\%	&37.0\%	&41.1\%	&39.4\%	&42.4\%	&36.7\%\\
        {Harp} & 53.2\%&	59.7\%&	59.4\%&	59.7\%&	60.4\%&	54.2\%&	60.3\%&	53.8\%\\
        \bottomrule
    \end{tabular}
\end{table*}

\subsection{Reduction to a Two-Stage Subproblem}
We employ mathematical induction to extend the two-stage base case to a pipeline with $S$ stages. Assume that for a pipeline with $S{-}1$ stages (from stage 2 to $S$), the H-1F1B scheduler already achieves the minimum steady-phase latency and a bubble-free execution, where each stage $i \in \{2, \dots, S{-}1\}$ satisfies the launch count $N_i$. 

The key observation is that the start times of all operations on stages $3$ through $S$ are recursively determined by the schedule of stage 2. Specifically, if the relative start times of nodes on stage 2 remain consistent with the $(S{-}1)$-stage optimal schedule, the later stages will remain bubble-free. As shown in Fig.~\ref{fig:pipeline_dag}(c), by pruning nodes and edges unrelated to the boundary between stage 1 and stage 2, we can model the interaction as an equivalent two-stage system where the schedule for stages $2 \dots S$ is fixed.

\subsection{Path Analysis and Precedence Constraints}

In this reduced setting, we analyze the temporal distance between the forward pass $F_{i,1}$ and the backward pass $B_{i,1}$ for any microbatch $i$ in the steady phase. We categorize the paths from $F_{i,1}$ to $B_{i,1}$ into two types:

\noindent\textbf{(1) Local Path ($\gamma = 0$):} A path consisting solely of operations within stage 1. Due to the $N_1$ microbatches launched during the warm-up phase, the interval between $F_{i,1}$ and $B_{i,1}$ must accommodate $N_1$ forward and $N_1-1$ backward microbatch computations, resulting in a total path length of $N_1 f + (N_1-1)b$.

\noindent\textbf{(2) Round-Trip Paths ($\gamma \ge 1$):} 
These paths represent dependencies that traverse the interconnect between stage 1 and stage 2. Specifically, $\gamma$ denotes the number of round-trips where a dependency chain flows through stage 2 and returns to stage 1. Generalizing from the two-stage case, the length of such a path is given by:
\begin{equation}
[N_2+(\gamma-1)(1-\delta_1)+1]f + [N_2+(\gamma-1)(1-\delta_1)]b + 2c_1\gamma.
\end{equation}

Notably, the Local Path is a special case of this formulation where $\gamma = 0$. By incorporating all possible values of $\gamma$, the temporal distance between the start times of $F_{i,1}$ and $B_{i,1}$ is constrained by the maximum duration across all potential dependency paths:
\begin{align}
s(B_{i,1}) = s(F_{i,1}) + \max_{\gamma \ge 0} \{ &[N_2+(\gamma-1)(1-\delta_1)+1]f + \\
&[N_2+(\gamma-1)(1-\delta_1)]b + 2c_1\gamma \}.\notag
\end{align}

Similarly, the time required to complete a "block" of $N_1$ microbatches is:
\begin{align}
    s(F_{i+N_1,1}) =& s(F_{i,1}) + \max_\gamma \{ 2c_1\gamma  + \\\notag
    &[N_2+(\gamma-1)(1-\delta_1)+1]\cdot(f+b) \}.
\end{align}

\subsection{End-to-End Latency Analysis}
Given a total of $B$ microbatches, the number of steady-phase execution blocks is approximately $(B - N_1) / N_1$. For a sufficiently large $B$, the total steady-phase duration $T(\delta_1)$ is dominated by the following expression:
\begin{align}
   T(\delta_1) &\;=\; O(1) + \\
&B \cdot \max_\gamma\!\Big\{\frac{[N_2+(\gamma-1)(1-\delta_1)+1]\cdot (f+b)+2c_1\gamma}{N_2+\delta_1}\Big\}\,,\notag 
\end{align}
where $O(1)$ represents the overhead from the pipeline warm-up and cool-down phases. Let $z(\delta_1, \gamma)$ denote the per-microbatch latency term within the $\max$ operator:
\begin{align}
    z(\delta_1,\gamma) = \frac{[N_2+(\gamma-1)(1-\delta_1)+1]\cdot (f+b)+2c_1\gamma}{N_2+\delta_1}.
\end{align}
We observe that for the local path ($\gamma=0$), the expression simplifies to $z(\delta_1, 0) = f+b$, which represents the theoretical minimum latency (the "bubble-free" case). Since the local path always exists as a lower bound, achieving optimal throughput requires that no round-trip path ($\gamma \ge 1$) exceeds the duration of the local path. This leads to the following optimality condition:
\begin{equation}
    z(\delta_1,\gamma) \le f+b.
\end{equation}

By substituting the definition of $z(\delta_1, \gamma)$ and simplifying the inequality, we can eliminate the $\gamma$, yielding:
\begin{equation}
    \delta_1\ge\frac{2c}{f+b}+1.
\label{eq:s_stage_delta_final}
\end{equation}
Under the H-1F1B scheduling rule, $\delta_1$ is selected as the smallest integer satisfying Eq.~\eqref{eq:s_stage_delta_final}, thus ensuring the theoretical upper bound of communication hiding is achieved. This completes the proof.

%% file: appendix/B-MFU.tex
\section{MFU and parallel strategy of End-to-End Experiments.}
\label{MFU}

\begin{table}[H]
    \caption{Parallel Strategy of Harp in Figure-\ref{fig:evaluation_e2e_heter}(a), MFU=53.2\%.}
    \centering
    \small
    \begin{tabular}{cccccc}
        \toprule
            \makecell{\textbf{Stage}} &
            \makecell{\textbf{Layer}\\\textbf{Range}} &
            \makecell{\textbf{Cluster}} &
            \makecell{\textbf{Physical}\\\textbf{Mesh}} &
            \makecell{\textbf{Logical}\\\textbf{Mesh}} &
            \makecell{\textbf{Force DP}\\\textbf{Batch Dim}} \\
        \midrule
            1 & [0, 39]    & V100e & (1, 8) & (2, 4) & 0 \\
        \midrule
            2 & [40, 92]   & A100  & (1, 4) & (1, 4) & 0 \\
            3 & [93, 145]  & A100  & (1, 4) & (1, 4) & 0 \\
        \bottomrule
    \end{tabular}
\end{table}

\begin{table}[H]
    \caption{Parallel Strategy of Harp in Figure-\ref{fig:evaluation_e2e_heter}(b), MFU=59.7\%.}
    \centering
    \small
    \begin{tabular}{cccccc}
        \toprule
            \makecell{\textbf{Stage}} &
            \makecell{\textbf{Layer}\\\textbf{Range}} &
            \makecell{\textbf{Cluster}} &
            \makecell{\textbf{Physical}\\\textbf{Mesh}} &
            \makecell{\textbf{Logical}\\\textbf{Mesh}} &
            \makecell{\textbf{Force DP}\\\textbf{Batch Dim}} \\
        \midrule
            1 & [0, 15]   & V100e & (1, 4) & (4, 1) & 0 \\
            2 & [16, 32]  & V100e & (1, 4) & (4, 1) & 0 \\
            3 & [33, 49]  & V100e & (1, 4) & (4, 1) & 0 \\
            4 & [50, 66]  & V100e & (1, 4) & (4, 1) & 0 \\
        \midrule
            5 & [67, 76]  & A100 & (1, 1) & (1, 1) & 0 \\
            6 & [77, 86]  & A100 & (1, 1)  & (1, 1) & 0 \\
            7 & [87, 106] & A100 & (1, 2)  & (2, 1) & 0 \\
            8 & [107, 126]& A100 & (1, 2) & (2, 1) & 0 \\
            9 & [127, 145]& A100 & (1, 2) & (2, 1) & 0 \\
        \bottomrule
    \end{tabular}
\end{table}

\begin{table}[H]
    \caption{Parallel Strategy of Harp in Figure-\ref{fig:evaluation_e2e_heter}(c), MFU=59.4\%.}
    \centering
    \small
    \begin{tabular}{cccccc}
        \toprule
            \makecell{\textbf{Stage}} &
            \makecell{\textbf{Layer}\\\textbf{Range}} &
            \makecell{\textbf{Cluster}} &
            \makecell{\textbf{Physical}\\\textbf{Mesh}} &
            \makecell{\textbf{Logical}\\\textbf{Mesh}} &
            \makecell{\textbf{Force DP}\\\textbf{Batch Dim}} \\
        \midrule
            1 & [0, 12]   & V100e & (1, 4) & (4, 1) & 0 \\
            2 & [13, 26]  & V100e & (1, 4) & (4, 1) & 0 \\
            3 & [27, 40]  & V100e & (1, 4) & (4, 1) & 0 \\
            4 & [41, 54]  & V100e & (1, 4) & (4, 1) & 0 \\
            5 & [55, 82]  & V100e & (1, 8) & (8, 1) & 0 \\
        \midrule
            6 & [83, 97]  & A100  & (1, 2) & (2, 1) & 0 \\
            7 & [98, 114] & A100  & (1, 2) & (2, 1) & 0 \\
            8 & [115, 145]& A100  & (1, 4) & (4, 1) & 0 \\
        \bottomrule
    \end{tabular}
\end{table}

\begin{table}[H]
    \caption{Parallel Strategy of Harp in Figure-\ref{fig:evaluation_e2e_heter}(d), MFU=59.7\%.}
    \centering
    \small
    \begin{tabular}{cccccc}
        \toprule
            \makecell{\textbf{Stage}} &
            \makecell{\textbf{Layer}\\\textbf{Range}} &
            \makecell{\textbf{Cluster}} &
            \makecell{\textbf{Physical}\\\textbf{Mesh}} &
            \makecell{\textbf{Logical}\\\textbf{Mesh}} &
            \makecell{\textbf{Force DP}\\\textbf{Batch Dim}} \\
        \midrule
            1 & [0, 21]    & V100e & (1, 8) & (8, 1) & 0 \\
            2 & [22, 44]   & V100e & (1, 8) & (8, 1) & 0 \\
            3 & [45, 67]   & V100e & (1, 8) & (8, 1) & 0 \\
            4 & [68, 90]   & V100e & (1, 8) & (8, 1) & 0 \\
        \midrule
            5 & [91, 118]  & A100  & (1, 4) & (4, 1) & 0 \\
            6 & [119, 145] & A100  & (1, 4) & (4, 1) & 0 \\
        \bottomrule
    \end{tabular}
\end{table}

\begin{table}[H]
    \caption{Parallel Strategy of Harp in Figure-\ref{fig:evaluation_e2e_heter}(e), MFU=60.4\%.}
    \centering
    \small
    \begin{tabular}{cccccc}
        \toprule
            \makecell{\textbf{Stage}} &
            \makecell{\textbf{Layer}\\\textbf{Range}} &
            \makecell{\textbf{Cluster}} &
            \makecell{\textbf{Physical}\\\textbf{Mesh}} &
            \makecell{\textbf{Logical}\\\textbf{Mesh}} &
            \makecell{\textbf{Force DP}\\\textbf{Batch Dim}} \\
        \midrule
            1  & [0, 10]    & V100e & (1, 4) & (4, 1) & 0 \\
            2  & [11, 22]   & V100e & (1, 4) & (4, 1) & 0 \\
            3  & [23, 34]   & V100e & (1, 4) & (4, 1) & 0 \\
            4  & [35, 46]   & V100e & (1, 4) & (4, 1) & 0 \\
            5  & [47, 58]   & V100e & (1, 4) & (4, 1) & 0 \\
            6  & [59, 70]   & V100e & (1, 4) & (4, 1) & 0 \\
            7  & [71, 82]   & V100e & (1, 4) & (4, 1) & 0 \\
            8  & [83, 94]   & V100e & (1, 4) & (4, 1) & 0 \\
        \midrule
            9  & [95, 107]  & A100  & (1, 2) & (1, 2) & 0 \\
            10 & [108, 132] & A100  & (1, 4) & (2, 2) & 0 \\
            11 & [133, 145] & A100  & (1, 2) & (2, 1) & 0 \\
        \bottomrule
    \end{tabular}
\end{table}

\begin{table}[H]
    \caption{Parallel Strategy of Harp in Figure-\ref{fig:evaluation_e2e_heter}(f), MFU=54.2\%.}
    \centering
    \small
    \begin{tabular}{cccccc}
        \toprule
            \makecell{\textbf{Stage}} &
            \makecell{\textbf{Layer}\\\textbf{Range}} &
            \makecell{\textbf{Cluster}} &
            \makecell{\textbf{Physical}\\\textbf{Mesh}} &
            \makecell{\textbf{Logical}\\\textbf{Mesh}} &
            \makecell{\textbf{Force DP}\\\textbf{Batch Dim}} \\
        \midrule
            1  & [0, 9]     & V100e & (1, 4) & (2, 2) & 0 \\
            2  & [10, 19]   & V100e & (1, 4) & (2, 2) & 0 \\
            3  & [20, 29]   & V100e & (1, 4) & (2, 2) & 0 \\
            4  & [30, 39]   & V100e & (1, 4) & (2, 2) & 0 \\
        \midrule
            5  & [40, 46]   & A100  & (1, 1) & (1, 1) & 0 \\
            6  & [47, 53]   & A100  & (1, 1) & (1, 1) & 0 \\
            7  & [54, 60]   & A100  & (1, 1) & (1, 1) & 0 \\
            8  & [61, 66]   & A100  & (1, 1) & (1, 1) & 0 \\
            9  & [67, 73]   & A100  & (1, 1) & (1, 1) & 0 \\
            10 & [74, 80]   & A100  & (1, 1) & (1, 1) & 0 \\
            11 & [81, 93]   & A100  & (1, 2) & (2, 1) & 0 \\
            12 & [94, 106]  & A100  & (1, 2) & (2, 1) & 0 \\
            13 & [107, 119] & A100  & (1, 2) & (2, 1) & 0 \\
            14 & [120, 132] & A100  & (1, 2) & (2, 1) & 0 \\
            15 & [133, 145] & A100  & (1, 2) & (2, 1) & 0 \\
        \bottomrule
    \end{tabular}
\end{table}

\begin{table}[H]
    \caption{Parallel Strategy of Harp in Figure-\ref{fig:evaluation_e2e_heter}(g), MFU=60.3\%.}
    \centering
    \small
    \begin{tabular}{cccccc}
        \toprule
            \makecell{\textbf{Stage}} &
            \makecell{\textbf{Layer}\\\textbf{Range}} &
            \makecell{\textbf{Cluster}} &
            \makecell{\textbf{Physical}\\\textbf{Mesh}} &
            \makecell{\textbf{Logical}\\\textbf{Mesh}} &
            \makecell{\textbf{Force DP}\\\textbf{Batch Dim}} \\
        \midrule
            1  & [0, 15]    & V100e & (1, 8) & (4, 2) & 0 \\
            2  & [16, 24]   & V100e & (1, 4) & (4, 1) & 0 \\
            3  & [25, 33]   & V100e & (1, 4) & (4, 1) & 0 \\
            4  & [34, 42]   & V100e & (1, 4) & (4, 1) & 0 \\
            5  & [43, 51]   & V100e & (1, 4) & (4, 1) & 0 \\
            6  & [52, 60]   & V100e & (1, 4) & (4, 1) & 0 \\
            7  & [61, 69]   & V100e & (1, 4) & (4, 1) & 0 \\
        \midrule
            8  & [70, 79]   & A100  & (1, 2) & (1, 2) & 0 \\
            9  & [80, 89]   & A100  & (1, 2) & (1, 2) & 0 \\
            10 & [90, 108]  & A100  & (1, 4) & (2, 2) & 0 \\
            11 & [109, 127] & A100  & (1, 4) & (2, 2) & 0 \\
            12 & [128, 145] & A100  & (1, 4) & (2, 2) & 0 \\
        \bottomrule
    \end{tabular}
\end{table}

\begin{table}[H]
    \caption{Parallel Strategy of Harp in Figure-\ref{fig:evaluation_e2e_heter}(h), MFU=53.8\%.}
    \centering
    \small
    \begin{tabular}{cccccc}
        \toprule
            \makecell{\textbf{Stage}} &
            \makecell{\textbf{Layer}\\\textbf{Range}} &
            \makecell{\textbf{Cluster}} &
            \makecell{\textbf{Physical}\\\textbf{Mesh}} &
            \makecell{\textbf{Logical}\\\textbf{Mesh}} &
            \makecell{\textbf{Force DP}\\\textbf{Batch Dim}} \\
        \midrule
            1  & [0, 9]     & V100e & (1, 8) & (4, 2) & 0 \\
            2  & [10, 19]   & V100e & (1, 8) & (4, 2) & 0 \\
            3  & [20, 29]   & V100e & (1, 8) & (4, 2) & 0 \\
            4  & [30, 39]   & V100e & (1, 8) & (4, 2) & 0 \\
        \midrule
            5  & [40, 46]   & A100  & (1, 2) & (2, 1) & 0 \\
            6  & [47, 53]   & A100  & (1, 2) & (2, 1) & 0 \\
            7  & [54, 60]   & A100  & (1, 2) & (2, 1) & 0 \\
            8  & [61, 67]   & A100  & (1, 2) & (2, 1) & 0 \\
            9  & [68, 80]   & A100  & (1, 4) & (4, 1) & 0 \\
            10 & [81, 93]   & A100  & (1, 4) & (4, 1) & 0 \\
            11 & [94, 106]  & A100  & (1, 4) & (4, 1) & 0 \\
            12 & [107, 119] & A100  & (1, 4) & (4, 1) & 0 \\
            13 & [120, 132] & A100  & (1, 4) & (4, 1) & 0 \\
            14 & [133, 145] & A100  & (1, 4) & (4, 1) & 0 \\
        \bottomrule
    \end{tabular}
\end{table}

In Appendix \ref{MFU}, we provide the Model Flops Utilization (MFU) in Table\ref{tab:mfu} and detailed parallel strategies for the end-to-end experiments discussed in Section \ref{section:6.1}. In our evaluation, the V100 and V100e are emulated using NVIDIA A100 GPUs by throttling the GPU clock frequency via \texttt{nvidia-smi -lgc 585, 585}. While this clock frequency accurately emulates the GEMM computational performance of the V100, the I/O throughput remains higher than that of native V100 GPUs. Consequently, the MFU values reported in this study are slightly higher than those achievable on physical V100 hardware.

However, it is important to emphasize that this evaluation remains fair for two reasons: (1) \textsc{Harp} bases its optimization on empirical performance metrics obtained from the profiler rather than hard-coded hardware specifications, and (2) all baselines are evaluated using the same emulated environment. Therefore, the performance gains and comparative results presented in this work remain consistent and unbiased.

Besides, when analyzing the experimental results, we focus on vertical comparisons—evaluating the performance of different frameworks within the same heterogeneous setting. In contrast, horizontal comparisons of MFU across different heterogeneous settings are generally discouraged. This is because the MFU in a heterogeneous environment is derived as a weighted average of disparate hardware capacities, which can make such cross-setting comparisons susceptible to Simpson's Paradox. Consequently, a higher aggregate MFU in one heterogeneous configuration does not necessarily imply superior hardware utilization over a lower MFU in another.

%% file: appendix/C-HardwareSpec.tex
\section{Hardware specifications of Heterogeneous Setting 1}
\begin{table*}[ht]
    \centering
    \caption{Hardware specifications of Heterogeneous Setting 1 (BW: Bandwidth).}    
    \label{tab:gpu_specs_transposed}
    \begin{tabular}{lcccccc}
        \toprule
        & \textbf{Number} & \textbf{FP16 Tensor Core} & \textbf{GPU Memory} & \textbf{Intra-Host BW} & \textbf{Inter-Host BW} & \textbf{Cross-Cluster BW} \\
        \midrule
        \texttt{8$\times$A100} & 4 & 312 TFLOP/S & 40\,GB & 300\,GB/s & 200\,Gbps & 1-10\,Gbps \\
        \texttt{8$\times$V100e} & 4 & 125 TFLOP/S & 32\,GB & 150\,GB/s & 200\,Gbps & 1-10\,Gbps \\
        \bottomrule
    \end{tabular}
\end{table*}

Table~\ref{tab:gpu_specs_transposed} summarizes the detailed hardware specifications for Heterogeneous Setting 1.

%% file: appendix/D-evaluationMetis.tex
\section{Discussion on the Exclusion of Metis}

We do not include Metis~\cite{um2024metis} in our comparative analysis primarily due to its architectural constraints. First, the current implementation of the Metis planner requires a uniform number of GPUs across all nodes, including heterogeneous ones. This constraint is incompatible with many real-world heterogeneous scenarios, such as our Heterogeneous Setting 2 and 3, where enforcing such uniformity would restrict the usable GPU count to only 6 instead of the full 12, leading to significant throughput degradation. Second, prior evaluations by Yan et al.\cite{yan2024hexiscale} have demonstrated that HexiScale consistently outperforms Metis, achieving performance improvements of $1.6\times$ to $1.9\times$. Given that HexiScale already represents a state-of-the-art advancement over Metis, we focus our comparative analysis on HexiScale.